\documentclass[journal]{IEEEtran}

\usepackage{booktabs}
\usepackage{dsfont}
\usepackage{threeparttable}
\usepackage{amsmath,graphicx,cite,amssymb}
\usepackage{amsfonts,multirow,bm,array,setspace}

\usepackage{graphicx,float,cite,amssymb}
\usepackage{multirow,bm,array,setspace}
\usepackage{textcomp}
\usepackage{psfrag}
\usepackage{color}
\usepackage{pstricks}

\usepackage[linesnumbered,ruled]{algorithm2e}
\makeatletter
\newcommand{\distas}[1]{\mathbin{\overset{#1}{\kern\z@\sim}}}%
\newsavebox{\mybox}\newsavebox{\mysim}
\newcommand{\distras}[1]{%
  \savebox{\mybox}{\hbox{\kern1pt$\scriptstyle#1$\kern1pt}}%
  \savebox{\mysim}{\hbox{$\sim$}}%
  \mathbin{\overset{#1}{\kern\z@\resizebox{\wd\mybox}{\ht\mysim}{$\sim$}}}%
}
\makeatother
\ifCLASSINFOpdf
\else
\fi
\newtheorem{proposition}{Proposition}

\newtheorem{lemma}{Lemma}
\newtheorem{theorem}{Theorem}
\newtheorem{corollary}{Corollary}

\newcommand{\bA}{\mathbf A}
\newcommand{\ba}{\mathbf a}
\newcommand{\bB}{\mathbf B}
\newcommand{\be}{\mathbf e}

\newcommand{\bg}{\mathbf g}

\newcommand{\bE}{\mathbf E}

\newcommand{\bx}{\mathbf{x}}

\newcommand{\bI}{\bm{I}}

\newcommand{\bv}{\mathbf v}

\newcommand{\bU}{\mathbf U}
\newcommand{\bY}{\mathbf{Y}}

\newcommand{\bQ}{\mathbf{Q}}

\newcommand{\bR}{\mathbf{R}}

\newcommand{\bh}{\mathbf{h}}

\newcommand{\bW}{\mathbf{W}}
\newcommand{\bD}{\mathbf{D}}
\newcommand{\bz}{\mathbf{z}}

\newcommand{\bZ}{\mathbf{Z}}
\newcommand{\bbE}{\mathbb{E}}

\newcommand{\bphi}{\bm\phi}

\newcommand{\bpsi}{\bm\psi}
\newcommand{\bvphi}{\bm\varphi}
\newcommand{\bmu}{\bm\lambda}
\newcommand{\bLambda}{\bm\Lambda}

\begin{document}
%\title{Full-Duplex Cellular System with Cross-Channel Transmission and Interference Cancellation}
\title{Enhanced Channel Estimation in Massive MIMO via Coordinated Pilot Design}

\author{\IEEEauthorblockN{Kaiming Shen, \IEEEmembership{Member,~IEEE}, Hei Victor Cheng, \IEEEmembership{Member,~IEEE},\\
Xihan Chen, \IEEEmembership{Student Member,~IEEE}, Yonina C. Eldar, \IEEEmembership{Fellow,~IEEE}, and Wei Yu, \IEEEmembership{Fellow,~IEEE}} % <-this % stops a space
\thanks{Manuscript received December 18, 2019, revised June 15, 2020 and July 28, 2020. The work of K. Shen was supported in part by the National Key R\&D Program of China with grant No. 2018YFB1800800 and in part by the Natural Sciences and Engineering Research Council (NSERC) of Canada. The work of H. V. Cheng and W. Yu was supported by the NSERC CRD program and the Canada Research Chairs program. The work of Y. C. Eldar was supported in part by European Union's Horizon 2020 Research and Innovation Program under Grant 646804-ERC-COG-BNYQ, in part by Futurewei Technologies, and in part by the Air Force Office of Scientific Research under Grant FA9550-18-1-0208.
This article was presented in part in IEEE
International Conference on Acoustics, Speech, and Signal Processing (ICASSP),
May 2019, Brighton, UK \cite{shen_icassp19}. {\it (Corresponding author: Kaiming Shen.)}

K. Shen was with
The Edward S. Rogers Sr. Department of Electrical and Computer
Engineering, University of Toronto, Toronto, ON M5S 3G4, Canada. He is now with the School of Science and Engineering, The Chinese
University of Hong Kong (Shenzhen), Shenzhen 518172, China (e-mail: shenkaiming@cuhk.edu.cn)

H. V. Cheng and W. Yu are with
The Edward S. Rogers Sr. Department of Electrical and Computer
Engineering, University of Toronto, Toronto, ON M5S 3G4, Canada
(e-mail: hei.cheng@utoronto.ca, weiyu@ece.utoronto.ca).

X. Chen is with the College of
Information Science and Electronic Engineering, Zhejiang University,
Hangzhou 310027, China (e-mail: chenxihan@zju.edu.cn).

Y. C. Eldar is with the Faculty of Mathematics and Computer Science,  Weizmann institute of Science, Rehovot 7610001, Israel  (e-mail: yonina.eldar@weizmann.ac.il).
}}%


%Telephone: (800) 555--1212, Fax: (888) 555--1212}


\maketitle



\begin{abstract}
Pilot contamination is a limiting factor in multicell massive multiple-input multiple-output (MIMO) systems because it can severely impair channel estimation. Prior works have suggested coordinating pilot design across cells in order to reduce the channel estimation error caused by pilot contamination. In this paper, we propose a method for coordinated pilot design using fractional programming to minimize the weighted mean squared-error (MSE) in channel estimation. In particular, we apply the recently proposed quadratic transform to the MSE expression which allows the effect of pilot contamination to be decoupled. The resulting problem reformulation enables the pilots to be optimized in closed form if they can be designed arbitrarily. When the pilots are restricted to a given set of orthogonal sequences, pilot optimization reduces to an assignment problem which can be solved by weighted bipartite matching. Furthermore, {\color{black}we consider the max-min fairness of data rates with orthogonal pilots} and obtain an extension of the proposed method to correlated Rayleigh fading. Finally, simulations demonstrate the advantage of the proposed (orthogonal and nonorthogonal) pilot designs as compared with state-of-the-art methods in combating pilot contamination.% and data transmission.
\end{abstract}

\section{Introduction}

\IEEEPARstart{A}{cquisition} of channel state information (CSI) is crucial in massive multiple-input
multiple-output (MIMO) wireless networks. A main challenge in channel
estimation is that due to the limited coherence time, pilot sequences assigned to multiple users across multiple cells cannot all be orthogonal. The
nonorthogonality between the pilots, e.g., when the same set of pilots is reused across cells,
causes the channel estimation for one user to be affected by
the pilots of other users. This effect is referred to in the literature as pilot contamination
\cite{larsson_edfors_Tufvesson_marzetta_05,lulu_14}.

\begin{figure}[t]
\begin{minipage}[b]{0.48\linewidth}
\psfrag{1}[][l]{$R_2$}
\psfrag{2}[][]{$R_1$}
\psfrag{3}[][]{$R_3$}
  \centering
  \centerline{\includegraphics[width=4.05cm]{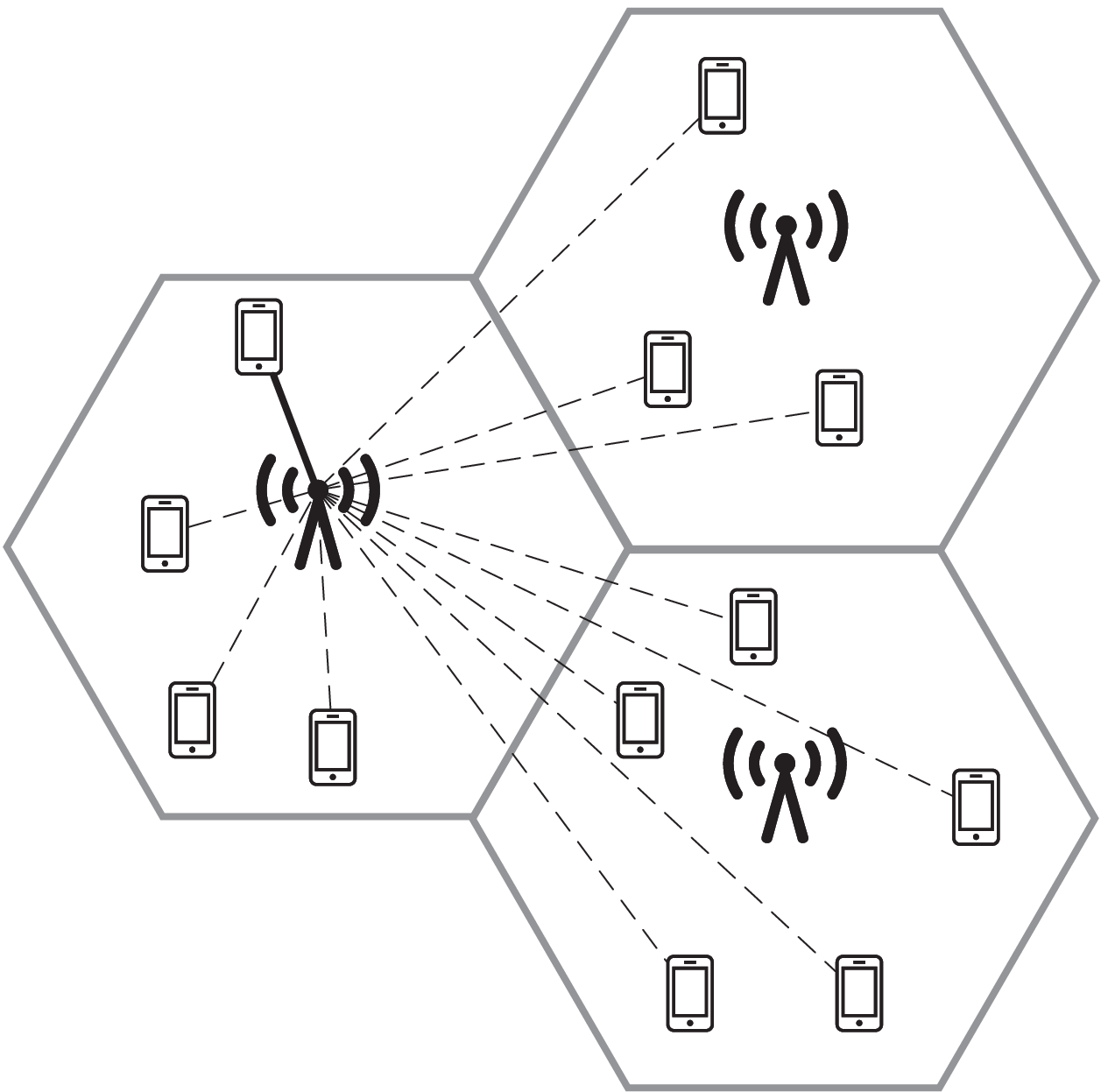}}
  \vspace{0.5em}
  \centerline{\footnotesize (a) Nonorthogonal pilots}\medskip
\end{minipage}
\hfill
\begin{minipage}[b]{0.5\linewidth}
\psfrag{1}[][l]{$R_2$}
\psfrag{2}[][]{$R_1$}
\psfrag{3}[][]{}
  \centering
  \centerline{\includegraphics[width=4.27cm]{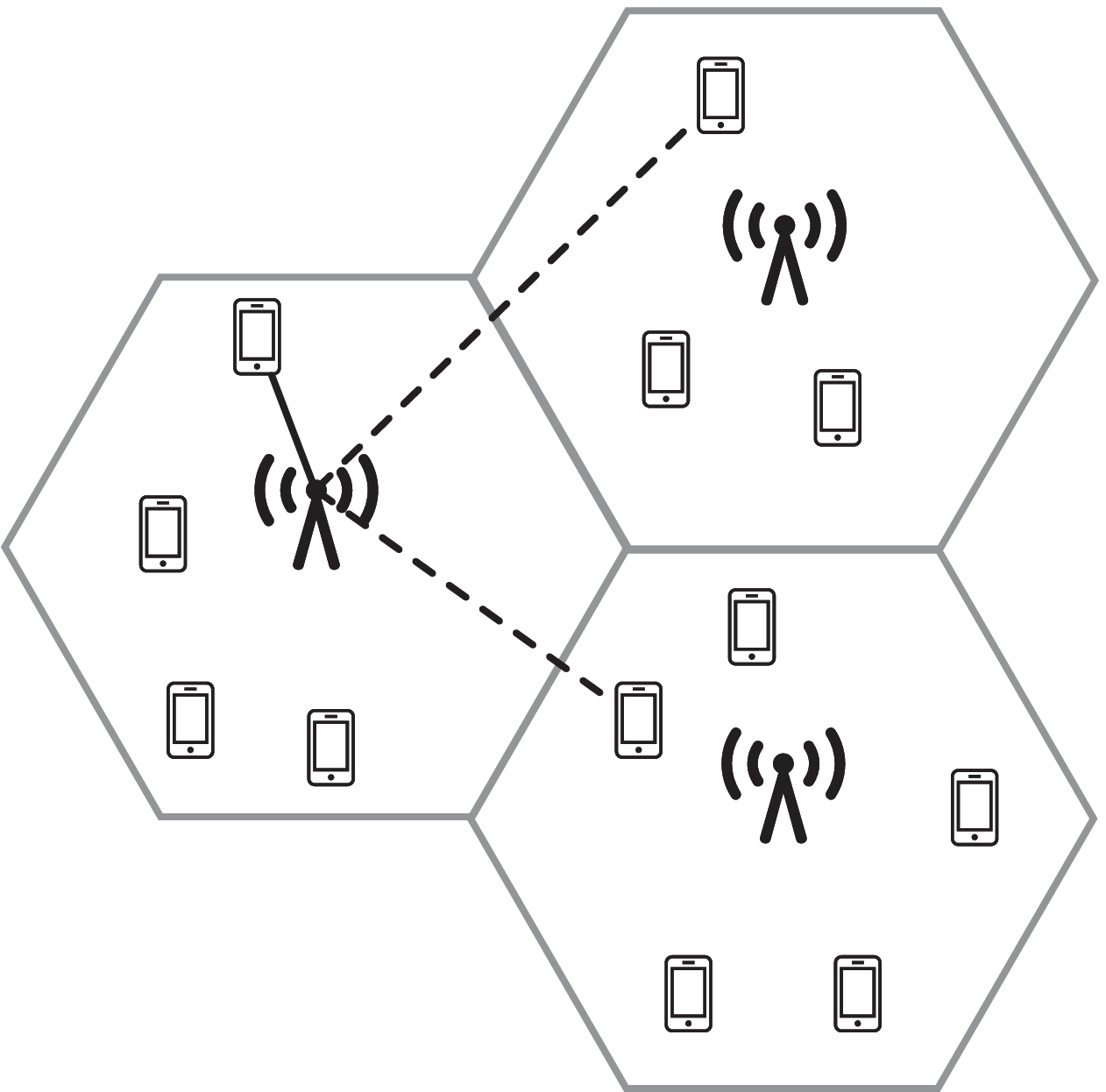}}
  \vspace{0.5em}
  \centerline{\footnotesize (b) Orthogonal pilots}\medskip
\end{minipage}
\caption{Orthogonal pilots versus nonorthogonal pilots. The solid line is the desired pilot while the dashed lines are the interfering pilots; the width of the dashed lines is proportional to the correlation with the desired pilot.}
\label{fig:pilot_contamination}
\end{figure}

This work pursues a strategy of designing pilot sequences of
user terminals across cells as a function of their large-scale fading (assuming that
user terminals are relatively stationary) in order to minimize pilot contamination.
Following the recent works of \cite{ioushua_eldar_17,eldar_18}, the idea is that the effect of
pilot contamination mainly depends on the large-scale fading between user terminals and
base stations (BSs). For example, if some interfering pilot signal is weak, then the desired pilots can afford to have higher correlation with it.
Thus, judicious pilot design for the different users across multiple cells
can help alleviate the pilot contamination effect.

The above goal can be further characterized as minimizing some suitable system-level metric of channel estimation performance by choosing the pilot sequences properly. The authors in \cite{ioushua_eldar_17,eldar_18} consider the minimum mean squared-error (MMSE) as the error metric. Here we additionally include weights, each reflecting the extent to which a particular user is affected by pilot contamination; thus weaker users may be assigned higher weights. We begin with the nonorthogonal case as illustrated in Fig.~\ref{fig:pilot_contamination}(a). Pilot design in this setting entails solving a multidimensional nonconvex problem. In contrast to standard tools such as greedy methods \cite{ioushua_eldar_17,eldar_18} and successive optimization \cite{salihi_nakhai_2018}, our approach is tailored to the fractional structure of the nonorthogonal pilot design problem. Specifically, minimizing the weighted MMSE using arbitrary pilots can be interpreted as a continuous sum-of-ratios programming problem. We simplify the problem by separating the numerator and denominator of each ratio. We achieve this separation by using the \emph{quadratic transform} \cite{part1,part2}, which is capable of decoupling more than one ratio. Earlier approaches to fractional programming (FP) such as the Dinkelbach's method \cite{dinkelbach_transform,crouzeix} cannot perform such a separation.

Although nonorthogonal pilots provide more accurate channel estimation, an orthogonal pilot scheme shown in Fig.~\ref{fig:pilot_contamination}(b) may still be favored in practice owing to its simple implementation. The assignment of orthogonal pilots to users involves a challenging combinatorial optimization.
In comparison to the state-of-the-art method \cite{zhu_wang_dai_qian_15} that assigns the orthogonal pilots to one cell at a time, we show that by using our decoupling approach, coordinated pilot design may be reformulated as a multi-cell assignment problem which can be efficiently solved via weighted bipartite matching.

%all the cells can be optimized simultaneously via weighted bipartite matching.% The advantage of this joint optimization is illustrated in our numerical results.

%The above results all rest on the assumption that channels are independently distributed. If correlated Rayleigh fading is added to the channel model, then the (weighted) MMSE channel estimation becomes a sum-of-matrix-ratios problem, wherein each mean squared-error (MSE) term is a trace of some matrix division. By applying matrix fractional programming as recently developed in \cite{matrix_FP}, we further generalize our pilot design to correlated Rayleigh fading.

{\color{black}
While \cite{muller_conttatellucci_vehkapera_14,ngo_larsson_icassp15,bose_17} consider blind channel estimation without using pilots, it is common in the literature of massive MIMO to enhance channel estimation accuracy via pilot optimization. The allocation of fixed orthogonal pilots across users is a well-studied problem in this area. In order to mitigate the pilot contamination effect, \cite{yin_gesbert_filippou_liu_13} proposes assigning pilots based on the orthogonality between the users' channels, which is quantified by the angle of arrival (AoA) of the received signal. Alternatively, \cite{zhu_wang_dai_qian_15} suggests a greedy method that optimizes the pilot assignment for one cell at a time. Another common heuristic \cite{su_yang_15,yan_yin_xia_wei_15} is to partially reuse pilots among cell-edge users; pilots are fully reused only among those center users that are more resistant to pilot contamination. In contrast to the above works assuming orthogonal pilots with fixed powers, a more sophisticated pilot design \cite{cheng_larsson_tsp2017,guo_icc14,liu_jiang_TCOM17,Hanzo_STSP14,Shariati_STSP14} incorporates power control into the orthogonal pilot assignment. As a further extension, \cite{chien_larsson2018} allows each user to combine multiple orthogonal pilots, but the resulting pilot sequences cannot be arbitrary as discussed in Section \ref{subsec:pilot_types}. Arbitrary nonorthogonal pilots are considered in \cite{ashikhmin_marzetta_12,ioushua_eldar_17,eldar_18}. A multi-cell precoding scheme is used in \cite{ashikhmin_marzetta_12} to combat pilot contamination, while \cite{ioushua_eldar_17,eldar_18} design the pilot symbols sequentially in a greedy fashion. In contrast to these existing works that consider only a particular type of pilot design, the FP-based framework proposed here can be used for both the orthogonal and nonorthogonal cases.

Prior works on pilot design can also be categorized according to their objective functions. The MSE-based metric has been considered extensively in the literature, including the sum-of-MSEs minimization \cite{yin_gesbert_filippou_liu_13,ioushua_eldar_17,eldar_18} and the sum-of-normalized-MSEs minimization \cite{liu_jiang_TCOM17,Hanzo_STSP14,Shariati_STSP14,bogale_le_15}. This work considers a general sum-of-weighted-MSEs minimization; the extension for correlated Rayleigh fading is also studied. Achievable rate is another common metric, e.g., \cite{cheng_larsson_tsp2017} considers maximizing an increasing concave utility function of rates, \cite{guo_icc14} aims to minimize the total power consumption under the rate constraint, and \cite{chien_larsson2018,zhu_wang_dai_qian_15} seek max-min fairness across user rates. In Section
\ref{subsec:maxmin} we show that max-min-rate optimization can be addressed from an FP perspective as well. We further provide a closed-form rate expression for arbitrary nonorthogonal pilots, whereas the result in \cite{chien_larsson2018} is for {\color{black}pilots which are nonnegative combinations of some fixed orthogonal sequences.}
}

The main contributions of this work include:
%\cite{ashikhmin_marzetta_12,ioushua_eldar_17,eldar_18}
{\color{black}
\begin{itemize}
\item \emph{Unified framework for pilot design:} The existing works mostly consider a particular type of pilot design. Based on our proposed approach of viewing weighted MSE minimization from an FP perspective, this work suggests a unified framework that accounts for both orthogonal pilot design and nonorthogonal pilot design.
\item \emph{Achievable rate analysis:} The achievable rate of massive MIMO is typically considered for orthogonal pilots; the recent work \cite{chien_larsson2018} gives an extension for a special type of {\color{black}nonorthogonal} pilots. This paper further generalizes the closed-form rate expression to arbitrary pilots.
\item \emph{Max-min fairness of user rates:} The well-known ``smart pilot assignment'' in \cite{zhu_wang_dai_qian_15} aims to maximize achievable rates with max-min fairness. We in addition consider the optimal pilot powers assuming that a given set of normalized orthogonal pilots have been assigned. It turns out that the power control problem, though nonconvex, can be efficiently solved via max-min-ratio FP.
\item \emph{Correlated channel case:} The MSE term has a matrix ratio form in the presence of correlated Rayleigh fading. We show that
    the numerator and denominator of each matrix ratio can still be decoupled by using a recent technique in \cite{matrix_FP}. As a result, the proposed FP framework for pilot design can be readily extended to the correlated channel case.
\end{itemize}
}

The rest of the paper is organized as follows. Section \ref{sec:system} describes the massive MIMO system and formulates the pilot design problem. Section \ref{sec:quadratic} briefly reviews the quadratic transform---a new FP technique \cite{part1,matrix_FP}. Section \ref{sec:nonorthogonal} examines the nonorthogonal pilot design while Section \ref{sec:orthogonal} treats the orthogonal setting. Section \ref{sec:correlated} extends the results to correlated channel estimation. Section \ref{sec:complexity} analyzes both computational complexity and communication complexity for the proposed algorithms. Numerical results are presented in Section \ref{sec:numerical}. Finally, Section \ref{sec:conclusion} concludes the paper.

Throughout the paper we use the following notation. We use $\|\cdot\|$ to denote the Euclidean norm, $(\cdot)^\top$ the
transpose, $(\cdot)^H$ the conjugate transpose, $\mathrm{vec}(\cdot)$ the vectorization, $\mathrm{tr}(\cdot)$
the trace. We let $\mathbb R$ be the set of real numbers, $\mathbb R_+$ the set of nonnegative numbers, $\mathbb C^{m\times n}$ the $m\times n$ dimensional complex space, $\mathbb H^{m\times m}$ the set of $m\times m$ Hermitian matrices. In addition, $\Re$ is the real
part of a complex number, $\bm I_n$ is an $n\times n$ identity matrix, $[1:n]$ is the discrete set $\{1,2,\ldots,n\}$, $\be^n_m$ is an $m\times1$ all-zeros vector except its $n$ entry being 1, and $\bE^{[n_1:n_2]}_m$ is an $m\times (n_2-n_1+1)$ matrix $[\be^{n_1}_m,\be^{n_1+1}_m,\ldots,\be^{n_2}_m]$. We use underline to denote a collection of variables, e.g., $\underline{\mathbf X}=\{{\mathbf X}_1,{\mathbf X}_2,\ldots,{\mathbf X}_n\}$. For ease of reference, we list the main variables in Table \ref{tab:notation}.

\begin{table}[t]
\small
\centering
\caption{\small List of Main Variables}
{\renewcommand{\arraystretch}{1.2}
\begin{tabular}{ll}
\hline
\hline
Notation &  Definition \\
\hline
$M$ & number of antennas at each BS\\
$L$ & number of cells\\
$K$ & number of user terminals per cell\\
$\tau$ & length of pilot\\
$l,i$  & index of BS or cell \\
$k,j$ & index of user terminal in the cell \\
$\bphi_{lk}$ & orthogonal or nonorthogonal pilot of user $(l,k)$ \\
$\bvphi_s$ & $s$th possible normalized orthogonal pilot\\
$\bpsi_{lk}$ & normalized orthogonal pilot assigned to user $(l,k)$\\
$p_{lk}$ & transmit power for $\bpsi_{lk}$ \\
$\beta_{ilk}$ & large-scale fading from user $(l,k)$ to BS $i$\\
$\bR_{ilk}$ & covariance matrix of from user $(l,k)$ to BS $i$\\
\hline
\hline
\end{tabular}}
\label{tab:notation}
\end{table}

\section{System Model}
\label{sec:system}

\subsection{Pilot Design Settings}
\label{subsec:pilot_types}

Consider an uplink massive MIMO system with $L$ cells, each cell consisting of one BS and $K$ user terminals. Assume that every BS has $M$ antennas and every user terminal has a single antenna. The full coherence bandwidth is reused across the cells. We use $(l,k)$ to index the $k$th user in the $l$th cell, for
$l\in[1:L]$ and $k\in[1:K]$; another index $(i,j)$ is similarly defined. Let $\bh_{lij}\in\mathbb C^M$ be the uplink channel from user $(i,j)$ to BS $l$. Each channel is modeled as
\begin{equation}
\label{chn}
\bh_{lij}=\sqrt{\beta_{lij}}\bg_{lij},
\end{equation}
in which the large-scale fading $\beta_{lij}$ is known \emph{a priori} while the Rayleigh fading $\bg_{lij}$ is drawn i.i.d. from a complex Gaussian distribution $\mathcal{CN}(\mathbf0,\bI_M)$. We begin with the above uncorrelated channel model. An extension with correlated $\bg_{lij}$ is provided in Section \ref{sec:correlated}.

Every pilot sequence consists of $\tau$ symbols. {\color{black}Assume that each channel $\bh_{lij}$ is fixed throughout the pilot sequence.} Let $\bphi_{lk}\in\mathbb C^\tau$ be the pilot sequence of user $(l,k)$. The received pilot signal $\bY_l\in\mathbb C^{M\times \tau}$ at BS $l$ can be expressed as
\begin{equation}
\bY_l = \sum_{(i,j)}\bh_{lij}\bphi_{ij}^\top +\bZ_l,
\end{equation}
where the additive background noise $\bZ_l\in\mathbb C^{M\times \tau}$ has each entry drawn i.i.d. from $\mathcal{CN}(0,\sigma^2)$. We compare three types of pilot design as follows:

\subsubsection{Orthogonal Pilots} Each pilot $\bphi_{lk}$ is structured as
\begin{equation}
\label{orthogonal_pilot}
\quad\bphi_{lk}=\sqrt{p_{lk}}\bpsi_{lk}\;\;\text{with}\;\;0<p_{lk}\le P_{\max},
\end{equation}
where $P_{\max}$ is the power constraint, and $\bpsi_{lk}$ is selected from a given set of normalized orthogonal pilots $\{\bvphi_1,\ldots,\bvphi_\tau\}$ with each $\|\bvphi_s\|^2=\tau$. In particular, the convention requires that the users in the same cell be assigned different pilots, e.g., $\bpsi_{lk}\ne\bpsi_{lk'}$ for $k\ne k'$. This type of pilot design is common in the existing literature.%This is the traditional pilot policy that has been extensively adopted in the prior works. %The optimization work here can be the power control $p_{lk}$ and the pilot assignment $\bpsi_{lk}$.
{\color{black}
\subsubsection{{\color{black}Restricted Nonorthogonal} Pilots}

The recent work \cite{chien_larsson2018} proposes that each user $(l,k)$ sends a nonnegative combination of some fixed orthogonal sequences:
{\color{black}
\begin{equation}
\label{semi_orthogonal}
\bphi_{lk}=\sum^\tau_{s=1}c^{(s)}_{lk}\bvphi_s\;\;\text{with}\;\;\sum^\tau_{s=1}\big|c^{(s)}_{lk}\big|^2\le P_{\max}\text{ and } c^{(s)}_{lk}\in\mathbb R_+.
\end{equation}}
Since each $c^{(s)}_{lk}$ is nonnegative, the choice of pilot is restricted to the positive orthant of the linear space spanned by the normalized orthogonal pilots $\{\bvphi_1,\ldots,\bvphi_\tau\}$. This assumption is critical to the geometric programming method in \cite{chien_larsson2018}.
}

\subsubsection{Nonorthogonal Pilots}
A further generalization allows each $\bphi_{lk}$ to be an arbitrary sequence in the $\tau$-dimensional space under the power constraint:
\begin{equation}
\label{nonorthogonal_pilot}
\bphi_{lk}\in\mathbb C^\tau \;\;\text{with}\;\;\|\bphi_{lk}\|^2\le \tau P_{\max}.
\end{equation}
This general form of pilots has been studied in \cite{ashikhmin_marzetta_12,ioushua_eldar_17,eldar_18}. {\color{black}Note that (\ref{semi_orthogonal}) is equivalent to (\ref{nonorthogonal_pilot}) when each $c^{(s)}_{lk}$ can be an arbitrary complex number not restricted to $\mathbb R_+$.}

%In addition, \cite{chien_larsson2018} considers a special type of nonorthogonal pilots structured as $\bphi_{lk}=\sum^\tau_{s=1}\sqrt{p^s_{lk}}\bvphi_s$, which corresponds to the positive orthant of the $\tau$-dimensional space with respect to the basis $\{\bvphi_1,\ldots,\bvphi_\tau\}$; this special assumption is critical to the geometric programming method in \cite{chien_larsson2018}.

% We remark that case with $\zeta_{lkt}$ in (\ref{nonorthogonal_structure}) restricted to the set of positive real numbers is studied in

\subsection{Weighted MSE Minimization}
Based on the received pilot signal $\bY_l$, each BS $l$ aims to recover its own channels $\{\bh_{ll1},\ldots,\bh_{llK}\}$. The channel estimate of $\bh_{llk}$ is chosen to minimize the MSE, i.e.,
\begin{equation}
\label{def_MMSE}
\hat\bh_{llk} = \arg\min_{\bh}\mathbb{E}\big[\|\bh_{llk}-\bh\|^2\big],
\end{equation}
where the expectation is taken over Rayleigh fading $\underline\bg$.
{\color{black}Following the standard steps as shown in \cite{ioushua_eldar_17,eldar_18}, we obtain the MMSE estimator at BS $l$:}
\begin{equation}
\label{H_est}
\hat\bh_{llk} = \big(\beta_{llk}\bphi^H_{lk}\otimes\bI_M\big)\big(\bD_l\otimes\bI_M\big)^{-1}\mathrm{vec}\big(\bY_l\big),
\end{equation}
where $\bD_l$ is the covariance matrix of $\bY_l$, i.e.,
\begin{equation}
\label{def_D}
{\color{black}\bD_{l} = \sigma^2\bm I_\tau+\sum_{(i,j)}\beta_{lij}\bphi_{ij}\bphi_{ij}^H.}
\end{equation}
The corresponding MSE of user $(l,k)$ is
\begin{equation}
\label{full_MSE}
\mathsf{MSE}_{lk} =M\beta_{llk}-M{\beta^2_{llk}}\Big(\bphi_{lk}^H\bD_l^{-1}\bphi_{lk}\Big).
\end{equation}
{\color{black}In contrast, the MMSE estimator of $\bh_{llk}$ in \cite{chien_larsson2018} is suboptimal because it is based on a partial projection $\bY_l\bphi_{lk}$ rather than the complete received signal $\bY_l$; it attains the minimum MSE in (\ref{full_MSE}) only when the orthogonal pilot design is assumed.}

%reduces to that of \cite{chien_larsson2018}

%Observe that the MSE is a function of the pilots $\underline\bphi$.

Given a set of positive weights $\alpha_{lk}>0$, we seek a set of pilots that lead to the minimum weighted sum MSE of channel estimation throughout the multicell system, i.e.,
\begin{equation}
\label{weighted_MSE}
\underset{\underline\bphi}{\text{minimize}}\quad\sum_{(l,k)}\alpha_{lk}\mathsf{MSE}_{lk}.
\end{equation}
The MSE weights $\alpha_{lk}$ are chosen on a case-by-case basis. For instance, we may set $\alpha_{lk}=1$ to minimize the sum of MSEs \cite{eldar_18}, or $\alpha_{lk}=1/\beta_{llk}$ to minimize the sum of normalized MSEs \cite{liu_jiang_TCOM17,Hanzo_STSP14,Shariati_STSP14,bogale_le_15}.% This work does not assume any particular choice of weights.

With (\ref{full_MSE}) substituted in (\ref{weighted_MSE}) and some constant terms removed, the above problem can be converted to
\begin{subequations}
\label{weighted_MSE_new}
\begin{align}
\underset{\underline\bphi}{\text{maximize}} &\quad
  \sum_{(l,k)}{\alpha_{lk}\beta^2_{llk}}\Big(\bphi_{lk}^H\bD_l^{-1}\bphi_{lk}\Big)
    \\
{\text{subject to}}&\quad \|\bphi_{lk}\|^2\le P_{\max}.
\end{align}
\end{subequations}
Here we assume that the pilots can be arbitrarily designed. % Observe that (\ref{weighted_MSE_new}) is a continuous nonconvex \bvphi_1 problem.
If an orthogonal pilot scheme is used, then an additional constraint (\ref{orthogonal_pilot}) is included in (\ref{weighted_MSE_new}). As a result, the problem involves the assignment of (normalized) orthogonal pilots $\{\bvphi_1,\ldots,\bvphi_\tau\}$ along with power control of $p_{lk}$.

\section{Quadratic Transform}
\label{sec:quadratic}

Our proposed approach to the pilot design problem in (\ref{weighted_MSE_new}) is based on the quadratic transform \cite{part1,matrix_FP}. This technique is capable of decoupling multiple (matrix) ratios simultaneously, whereas the traditional Dinkelbach's method \cite{dinkelbach_transform,crouzeix} is restricted to a single scalar ratio.

\begin{theorem}[Quadratic Transform \cite{part1}]
\label{theorem:FP}
Given a nonempty constraint set $\mathcal X$ as well as $N$ tuples of function $\mathbf a_n(\bx)\in\mathbb C^{m}$, function $\mathbf B_n(\bx)\in\mathbb H^{m\times m}$, and nondecreasing function $f_n:\mathbb R_+\mapsto\mathbb R$, for $n\in[1:N]$, the {sum-of-functions-of-ratio} problem
\begin{subequations}
\label{prob:FP}
\begin{align}
\underset{\bx}{\text{maximize}}\quad&
\sum^N_{n=1}f_n\Big({\mathbf a}^H_n(\bx)\mathbf B_n^{-1}(\bx){\mathbf a}_n(\bx)\Big)\\
\text{subject to}\quad& \bx\in\mathcal X
\end{align}
\end{subequations}
is equivalent to
\begin{subequations}
\label{prob:FP:fq}
\begin{align}
\underset{\bx,\,\underline\bmu}{\text{maximize}}\quad&
\sum^N_{n=1}f_n\Big(2\Re\{{\mathbf a}^H_n(\bx)\bmu_n\}
 - \bmu^H_n\mathbf B_n(\bx)\bmu_n \Big)\\
\text{subject to}\quad& \bx\in\mathcal X\\
\quad& \bmu_n\in\mathbb C^m,
\end{align}
\end{subequations}
where $\bmu_n$ is an auxiliary variable introduced for each ratio term ${\mathbf a}^H_n(\bx)\mathbf B_n^{-1}(\bx){\mathbf a}_n(\bx)$.
\end{theorem}

{\color{black}The equivalence between (\ref{prob:FP}) and (\ref{prob:FP:fq}) no longer holds when the argument of $f_n(\cdot)$ is a negative ratio. As shown in \cite{matrix_FP}, the quadratic transform amounts to constructing a \emph{surrogate function} so that the original objective function is lower bounded by the new objective function given any $\underline\bmu$; this can be guaranteed if and only if the value of each $\ba^H_n(\bx)\bB^{-1}_n(\bx)\ba_n(\bx)$ is positive.}

The quadratic transform can be further extended to the matrix ratio case as stated in the following theorem.

\begin{theorem}[Matrix Quadratic Transform \cite{matrix_FP}]
\label{theorem:FP_matrix}
Given a nonempty constraint set $\mathcal X$ as well as $N$ tuples of functions $\mathbf A_n(\bx)\in\mathbb C^{m_1\times m_2}$, functions $\mathbf B_n(\bx)\in\mathbb H^{m_1\times m_1}$, and nondecreasing functions $F_n:\mathbb{H}^{m_2\times m_2}\mapsto\mathbb R$ in the sense that $F_n(\mathbf C)\ge F_n(\mathbf C')$ if $\mathbf C\succeq\mathbf C'$, for $n\in[1:N]$, the {sum-of-functions-of-matrix-ratio} problem
\begin{subequations}
\label{prob:FP_matrix}
\begin{align}
\underset{\bx}{\text{maximize}}\quad&
\sum^N_{n=1}F_n\Big({\mathbf A}^H_n(\bx)\mathbf B_n^{-1}(\bx){\mathbf A}_n(\bx)\Big)\\
\text{subject to}\quad& \bx\in\mathcal X
\end{align}
\end{subequations}
is equivalent to
\begin{subequations}
\label{prob:FP_matrix:fq}
\begin{align}
\underset{\bx,\,\underline{\bLambda}}{\text{maximize}}\quad&
\sum^N_{n=1}F_n\Big(2\Re\{{\mathbf A}^H_n(\bx)\bLambda_n\}
 - \bLambda^H_n\mathbf B_n(\bx)\bLambda_n \Big)\\
\text{subject to}\quad& \bx\in\mathcal X\\
& \bLambda_n\in\mathbb C^{m_1\times m_2},
\end{align}
\end{subequations}
where $\bLambda_n$ is an auxiliary variable introduced for each matrix ratio term ${\mathbf A}^H_n(\bx)\mathbf B_n^{-1}(\bx){\mathbf A}_n(\bx)$.
\end{theorem}
%\begin{IEEEproof}
%By completing the square, each $\bY_i$ in (\ref{prob:quadratic_matrix:fq}) can be optimally computed as $\bY_i=\mathbf B^{-1}_i(\bx)\mathbf A_i(\bx)$. Plugging this optimal $\bY_i$ in (\ref{prob:quadratic_matrix:fq}) recovers (\ref{prob:quadratic_matrix}) and thus establishes the equivalence.
%\end{IEEEproof}

{\color{black}The quadratic transform in Theorem \ref{theorem:FP} suffices to deal with the problem in (\ref{weighted_MSE_new}), while its matrix version in Theorem \ref{theorem:FP_matrix} is used when we consider the correlated channel case in Section \ref{sec:correlated}.}
We then show that the (matrix) quadratic transform leads to an iterative optimization with provable convergence.

\begin{theorem}[Convergence Analysis \cite{matrix_FP}]
\label{theorem:convergence}
If $\bx$ and $\underline\bmu$ are optimized alternatingly in (\ref{prob:FP:fq}) or (\ref{prob:FP_matrix:fq}), the value of the original objective function is nondecreasing after each iteration. Furthermore, if the original objective function is differentiable, it converges to a stationary point of (\ref{prob:FP}) or (\ref{prob:FP_matrix}).
\end{theorem}

\begin{figure*}
\setcounter{equation}{23}
{\color{black}\begin{equation}
\label{rate}
R_{lk} = \log_2 \left(1+\frac{\beta^2_{llk}\big(\bphi^H_{lk}\bD^{-1}_l\bphi_{lk}\big)^2}{\frac{1}{M}\big(\sum_{(i,j)}\beta_{lij}+\sigma^2\big)\bphi^H_{lk}\bD^{-1}_l\bphi_{lk}+\sum_{(i,j)}\beta^2_{lij}\bphi^H_{lk}\bD^{-1}_l\bphi_{ij}\bphi_{ij}^H\bD^{-1}_l\bphi_{lk}-\beta^2_{llk}\big(\bphi^H_{lk}\bD^{-1}_l\bphi_{lk}\big)^2}\right).
\end{equation}
}
\hrule
\setcounter{equation}{15}
\end{figure*}

\section{Nonorthogonal Pilot Design}
\label{sec:nonorthogonal}

In this section we explore the use of the quadratic transform in nonorthogonal pilot design based on minimizing the MSE.
%Discussion on the orthogonal pilot design will be deferred until next section.
%\subsection{Iterative Optimization via Quadratic Transform}
The difficulty of problem (\ref{weighted_MSE_new}) lies in its fractional term $\bphi_{lk}^H\bD_l^{-1}\bphi_{lk}$, wherein the numerator and denominator are both affected by the pilot variable $\underline\bphi$. It is natural to decouple the numerator and denominator by using the quadratic transform in Theorem \ref{theorem:FP}. The resulting problem reformulation is stated in the following proposition.
\begin{proposition}
\label{prop:transform}
The nonorthogonal pilot design problem in (\ref{weighted_MSE_new}) is equivalent to
\begin{subequations}
\label{prob:quadratic}
\begin{align}
\underset{\underline\bphi,\,\underline\bmu}{\text{maximize}} &\quad
  f(\underline\bphi,\,\underline\bmu)
    \label{prob:quadratic:obj}\\
{\text{subject to}}&\quad \|\bphi_{lk}\|^2\le \tau P_{\max}
    \label{prob:quadratic:power}\\
&\quad \bmu_{lk}\in\mathbb C^{\tau},
\end{align}
\end{subequations}
where the new objective function is
\begin{equation}
\label{new_obj}
f(\underline\bphi,\,\underline\bmu)=\sum_{(l,k)}{\alpha_{lk}}\Big(2\beta_{llk}\Re\{\bmu_{lk}^H\bphi_{lk}\}-\bmu_{lk}^H\bD_l\bmu_{lk}\Big).
\end{equation}
\end{proposition}
\begin{IEEEproof}
The reformulation is obtained by treating $\beta_{llk}\bphi_{lk}$ and $\bD_l$ as $\mathbf a_n$ and $\bB_n$ in Theorem \ref{theorem:FP}, respectively, along with the nondecreasing function $f_n\big({\mathbf a}^H_n(\bx)\mathbf B_n^{-1}(\bx){\mathbf a}_n(\bx)\big)={\mathbf a}^H_n(\bx)\mathbf B_n^{-1}(\bx){\mathbf a}_n(\bx)$.
\end{IEEEproof}

We propose optimizing $\underline\bmu$ and $\underline\bphi$ alternatingly. As already shown in \cite{part1}, the auxiliary variable $\underline\bmu$ can be optimally updated by solving $\partial f/\partial \bmu_{lk}=\mathbf 0$ when $\underline\bphi$ is held fixed, resulting in
\begin{equation}
\label{opt_lambda}
\bmu^\star_{lk}= \beta_{llk}\bD_l^{-1}\bphi_{lk}.
\end{equation}
It remains to optimize the pilot variable $\underline\bphi$ for fixed $\underline\bmu$. It turns out that the solution can be obtained in closed form. To this end, we express $f(\underline\bphi,\,\underline\bmu)$ as
\begin{multline}
\label{rewrite_new_obj}
f(\underline\bphi,\,\underline\bmu) = \sum_{(l,k)}2\alpha_{lk} \beta_{llk}\Re\{\bmu^H_{lk}\bphi_{lk}\} \\ -\sum_{(l,k)}\bphi^H_{lk}\Bigg(\sum_{(i,j)} \alpha_{ij}\beta_{ilk}\bmu_{ij}\bmu^H_{ij}\Bigg)\bphi_{lk}+\mathrm{const},
\end{multline}
in which the last term $\mathrm{const}=\sum_{(l,k)}\alpha_{lk}\sigma^2\|\bmu^H_{lk}\|^2$ does not depend on $\underline\bphi$. The optimal pilots {\color{black}in terms of $f(\underline\bphi,\,\underline\bmu)$} are then easily solved, resulting in
\begin{equation}
\label{opt_phi}
\bphi_{lk}^\star = \Bigg(\sum_{(i,j)} \alpha_{ij}\beta_{ilk}\bmu_{ij}\bmu^H_{ij}+\eta_{lk}\bm I_\tau\Bigg)^{-1}\alpha_{lk} \beta_{llk}\bmu_{lk},
\end{equation}
where the Lagrange multiplier $\eta_{lk}$ accounts for the power constraint and is optimally determined as
\begin{equation}
\label{opt_eta}
	\eta^\star_{lk} =
\left\{
\begin{aligned}
&0,\;\text{if $\|\bphi^\star_{lk}\|^2\le \tau P_{\max}$ already};\\
&\text{$\eta_{lk}>0$ with $\|\bphi^\star_{lk}\|^2= \tau P_{\max}$, otherwise.}
\end{aligned}
\right.
\end{equation}
The evaluation of (\ref{opt_eta}) can be done by bisection search.

It can be readily obtained from Theorem \ref{theorem:convergence} that the iteration between (\ref{opt_lambda}) and (\ref{opt_phi}) leads to convergence.

\begin{proposition}
\label{prop:convergence}
The sum of weighted MSEs in (\ref{weighted_MSE}) is nonincreasing after each iteration in Algorithm \ref{alg:arbitrary_pilot}, while the pilot variable $\underline\bphi$ converges to a stationary point of the nonorthogonal pilot design problem in (\ref{weighted_MSE_new}).
\end{proposition}

To avoid the Lagrange multiplier $\eta_{lk}$, we take advantage of the observation in \cite{eldar_18} that multiplying all the pilots with the same nonzero scalar $\delta$ does not change the MSE values provided that the noise level $\sigma^2$ tends to zero. Thus, when the signal-to-noise ratio (SNR) is sufficiently high, we enforce the power constraint by scaling the pilots, without computing the Lagrange multiplier in (\ref{opt_eta}).

\begin{proposition}[Nonorthogonal Pilot Design Without Using Lagrange Multiplier]
\label{prop:fast_alg1}
If the noise level $\sigma^2\rightarrow0$, we can set $\eta_{lk}=0$ and determine $\bphi_{lk}$ as
\begin{equation}
\label{scale_phi}
\bphi^\star_{lk} = \delta \tilde{\bphi}_{lk},\;\text{for each}\; (l,k),
\end{equation}
where $\tilde{\bphi}_{lk}$ is obtained from (\ref{opt_phi}) with $\eta_{lk}=0$ and the scaling factor $\delta$ is computed as
\begin{equation}
\label{opt_delta}
	\delta = \min_{(l,k)}\frac{\sqrt{\tau P_{\max}}}{\|\tilde\bphi_{lk}\|}.
\end{equation}
The resulting $\underline\bphi^\star$ is a stationary point of the nonorthogonal pilot design problem in (\ref{weighted_MSE_new}).
\end{proposition}
\begin{IEEEproof}
For ease of discussion, we use (P1) to denote the original problem (\ref{weighted_MSE_new}), and (P2) the unconstrained version of (\ref{weighted_MSE_new}) with the power constraint removed. If $\underline\bphi'$ is a stationary point of (P2), then it is also a stationary point of (P1) so long as it meets the power constraint automatically.

According to Theorem \ref{theorem:convergence}, $\underline{\tilde\bphi}$ must be a stationary point of (P2). In addition, it can be shown that the first-order condition of (P2) remains the same after scaling every $\tilde\bphi_{lk}$ with $\delta$, so $\underline{\bphi}^\star$ must be a stationary point of (P2) as well. Note that $\underline{\bphi}^\star$ already meets the power constraint because of (\ref{opt_delta}), so it is also a stationary point of (P1).
\end{IEEEproof}

\SetEndCharOfAlgoLine{}
\begin{algorithm}[t]
Initialize the pilot variable $\underline\bphi$ to some feasible value\;
\Repeat{the weighted sum MSE converges\DontPrintSemicolon}{
    Update the auxiliary variable $\underline\bmu$ by (\ref{opt_lambda})\;
    \emph{Option 1:} Update the pilots $\underline\bphi$ by (\ref{opt_phi}) along with the Lagrangian multiplier $\eta_{lk}$ in (\ref{opt_eta})\;
    \emph{Option 2 (when $\sigma^2\rightarrow0$):} Update $\underline\bphi$ by (\ref{opt_phi}) with $\eta_{lk}=0$, then scale it as in (\ref{scale_phi}) and (\ref{opt_delta})\;
}
\caption{Proposed nonorthogonal pilot design for weighted MSE minimization}
\label{alg:arbitrary_pilot}
\end{algorithm}

Algorithm \ref{alg:arbitrary_pilot} summarizes the main procedure of the proposed nonorthogonal pilot design.
{\color{black}Next we discuss the resulting achievable rate. In the existing literature, the achievable rate of massive MIMO systems is typically considered for the orthogonal pilot case. Progress has been made in the recent work \cite{chien_larsson2018} to take {\color{black}a special type of nonorthogonal} pilots into account. The following theorem provides a further generalization that holds for arbitrary pilots.%, proof relegated to Appendix \ref{appendix:A}.

{\color{black}
\begin{theorem}[Achievable Rate with Nonorthogonal Pilots]
\label{theorem:rate}
Given a set of nonorthogonal pilots $\underline\bphi$ in (\ref{nonorthogonal_pilot}), the data rate $R_{lk}$ in (\ref{rate}) is achievable for user $(l,k)$.
\end{theorem}}
\begin{IEEEproof}
See Appendix \ref{appendix:A}.
\end{IEEEproof}
}

\section{Orthogonal Pilot Design}
\label{sec:orthogonal}

We now assume orthogonal pilots by imposing the constraint (\ref{orthogonal_pilot}) on the weighted MMSE problem (\ref{weighted_MSE_new}). With each $\bphi_{lk}$ expressed as $(p_{lk},\bpsi_{lk})$, the orthogonal pilot design problem can be formulated as
\setcounter{equation}{24}
\begin{subequations}
\label{weighted_MSE_new_orthogonal}
\begin{align}
\underset{\underline p,\,\underline\bpsi}{\text{maximize}} &\quad
  \sum_{(l,k)}{\alpha_{lk}\beta^2_{llk}}p_{lk}\Big(\bpsi_{lk}^H\bD_l^{-1}\bpsi_{lk}\Big)
    \\
{\text{subject to}}&\quad 0\le p_{lk}\le P_{\max}
    \label{weighted_MSE_new_orthogonal:b}\\
&\quad \bpsi_{lk}\in\{\bvphi_1,\ldots,\bvphi_\tau\}\\
&\quad\bpsi_{lk}\ne\bpsi_{lk'},\;\text{for any}\; k\ne k',
    \label{weighted_MSE_new_orthogonal:d}
\end{align}
\end{subequations}
where the covariance matrix $\bD_l$ of $\bY_l$ becomes
\begin{equation}
\label{def_D_new}
\bD_{l} = \sigma^2\bm I_\tau+\sum_{(i,j)}\beta_{lij}p_{ij}\bpsi_{ij}\bpsi_{ij}^H.
\end{equation}
The above problem has a mixed discrete-continuous form since it involves continuous variable $\underline p$ and discrete variable $\underline\bpsi$.

{\color{black}
\subsection{Orthogonal Pilot Assignment via Power Control}

The mixed discrete-continuous problem in (\ref{weighted_MSE_new_orthogonal}) is difficult to tackle directly. A naive idea is to reformulate it as a continuous power control problem. Specifically, introducing a new power variable $\tilde p^{(s)}_{lk}$ for each user $(l,k)$ and each possible orthogonal pilot $\bvphi_s$, we optimize the new power variable $\underline{\tilde p}$ and then assign some $\bvphi_s$ with nonzero $\tilde p^{(s)}_{lk}$ to each user $(l,k)$.

However, the resulting problem is still difficult because of the orthogonal pilot constraint. Since every user can choose only one pilot in $\{\bvphi_1,\ldots,\bvphi_\tau\}$, it requires that
\begin{align}
\label{naive_constr1}
    \big\|\tilde p^{(1)}_{lk},\ldots,\tilde p^{(\tau)}_{lk}\big\|_0&=1,\;l\in[1:L],k\in[1:K].
\end{align}
Moreover, if the users in the same cell cannot choose the same $\bvphi_s$, then we further have the following constraint:
\begin{align}
\label{naive_constr2}
    \tilde p^{(s)}_{lk}\tilde p^{(s)}_{lk'}&=0,\;l\in[1:L],k\ne k'\in[1:K],s\in[1:\tau].
\end{align}
Both (\ref{naive_constr1}) and (\ref{naive_constr2}) are difficult to handle.

Scalability is another issue since the new power variable $\tilde p^{(s)}_{lk}$ needs to be coordinated not only across users, but across all possible pilots $\bm\varphi_{s}$. Thus, rewriting (\ref{weighted_MSE_new_orthogonal}) in a continuous form does not necessarily make the problem easier. The rest of this section shows that the mixed discrete-continuous problem in (\ref{weighted_MSE_new_orthogonal}) can be efficiently addressed by means of weighted bipartite matching after the quadratic transform.
}

\subsection{Ratio Decoupling in Orthogonal Pilot Case}

\begin{algorithm}[t]
Initialize $(\underline p,\underline\bpsi)$ to some feasible point\;
\Repeat{the value of $\sum_{(l,k)}\alpha_{lk}\mathsf{MSE}_{lk}$ converges}{
    Update the auxiliary variable $\underline\bmu$ by (\ref{opt_lambda})\;
    \emph{Option 1:} Update $(\underline p,\underline\bpsi)$ by solving the weighted bipartite matching problem in (\ref{prob:matching})\;
    \emph{Option 2 (when constraint (\ref{weighted_MSE_new_orthogonal:d}) is removed):} Update $(\underline p,\underline\bpsi)$ by the linear search in (\ref{linear_search})\;
}
\caption{Proposed orthogonal pilot design for weighted MSE minimization}
\label{alg:orthogonal_pilot}
\end{algorithm}

The quadratic transform \cite{part1} still works in spite of the above changes. Following Proposition \ref{prop:transform}, we recast problem (\ref{weighted_MSE_new_orthogonal}) as
\begin{subequations}
\label{prob:quadratic_orthogonal}
\begin{align}
\underset{\underline p,\,\underline\bpsi,\,\underline\bmu}{\text{maximize}} &\quad
  f(\underline p,\underline\bpsi,\,\underline\bmu)
    \label{prob:quadratic_orthogonal:obj}\\
{\text{subject to}}& \quad \text{(\ref{weighted_MSE_new_orthogonal:b})--(\ref{weighted_MSE_new_orthogonal:d})}\\
&\quad \bmu_{lk}\in\mathbb C^{\tau},
    \label{prob:quadratic_orthogonal:e}
\end{align}
\end{subequations}
in which the new objective function is given by
\begin{multline}
\label{rewrite_new_obj}
f(\underline p,\underline\bpsi,\,\underline\bmu) = \sum_{(l,k)}2\sqrt{p_{lk}}\alpha_{lk} \beta_{llk}\Re\{\bmu^H_{lk}\bpsi_{lk}\} \\ -\sum_{(l,k)}p_{lk}\bpsi^H_{lk}\Bigg(\sum_{(i,j)} \alpha_{ij}\beta_{ilk}\bmu_{ij}\bmu^H_{ij}\Bigg)\bpsi_{lk}+\mathrm{const},
\end{multline}
where $\mathrm{const}$ refers to terms not depending on $(\underline p,\underline\bpsi)$.

As before, we propose to optimize the original variable $(\underline p,\underline\bpsi)$ and the auxiliary variable $\underline\bmu$ in an iterative fashion. When $(\underline p,\underline\bpsi)$ are held fixed, the optimal $\underline\bmu$ is still determined by (\ref{opt_lambda}) except that $\underline\bphi$ is replaced with $(\underline p,\underline\bpsi)$. In contrast, the optimization of pilots under fixed $\underline\bmu$ is quite different from the nonorthogonal case discussed in the previous section.

The key observation is that due to the convexity of (\ref{rewrite_new_obj}), the power variable $p_{lk}$ of user $(l,k)$ can be optimally determined for the new objective function $f(\underline p,\underline\bpsi,\,\underline\bmu)$ by solving the first-order equation $\partial f/\partial p_{lk}=0$, so long as the corresponding normalized sequence $\bpsi_{lk}$ is fixed. Hence, assuming that $\bpsi_{lk}=\bvphi_s$, for some $s\in[1:\tau]$, the optimal $p_{lk}$ {\color{black}in terms of $f(\underline p,\underline\bpsi,\,\underline\bmu)$} can be computed as
\begin{equation}
p_{lk}^{(s)} = \min\Bigg\{P_{\max},\,\Bigg(\frac{\alpha_{lk} \beta_{llk}\Re\{\bmu^H_{lk}\bvphi_{s}\}}{\bvphi_s^H\big(\sum_{(i,j)} \alpha_{ij}\beta_{ilk}\bmu_{ij}\bmu^H_{ij}\big)\bvphi_s}\Bigg)^2\Bigg\}.
\end{equation}

\subsection{Orthogonal Pilot Design via Weighted Bipartite Matching}
The new objective function $f$ in (\ref{rewrite_new_obj}) plays a crucial role in allowing each $p_{lk}$ to be optimized separately. Otherwise, the optimal $p_{lk}$ {\color{black}in terms of $f(\underline p,\underline\bpsi,\,\underline\bmu)$} would depend on the other variables $p_{ij}$ and $\bpsi_{ij}$ as in the original problem. Given $\bpsi_{lk}=\bvphi_s$, the tentative contribution of user $(l,k)$ to $f(\underline p,\underline\bpsi,\,\underline\bmu)$ is
\begin{multline}
\label{matching_weight}
\pi^{(s)}_{lk}=2\sqrt{p^{(s)}_{lk}}\alpha_{lk} \beta_{llk}\Re\{\bmu^H_{lk}\bvphi_{s}\}\\ -p^{(s)}_{lk}\bvphi^H_{s}\Bigg(\sum_{(i,j)} \alpha_{ij}\beta_{ilk}\bmu_{ij}\bmu^H_{ij}\Bigg)\bvphi_{s}.
\end{multline}
As a result, the maximization of $f(\underline p,\underline\bpsi,\,\underline\bmu)$ boils down to finding the optimal pair $(\bvphi_s,p_{lk}^{(s)})$ for each individual user, recognized as a weighted bipartite matching problem
\begin{subequations}
\label{prob:matching}
\begin{align}
\underset{\underline x}{\text{maximize}} &\quad
\sum_{(l,k,s)}\pi^{(s)}_{lk}x^{(s)}_{lk}
    \label{matching:obj}\\
{\text{subject to}}&\quad \sum^\tau_{s=1}x^{(s)}_{lk}=1,\;\text{for each}\; (l,k)
    \label{matching:sum_t_x}\\
&\quad \sum^K_{k=1}x^{(s)}_{lk}\le1,\;\text{for each}\; (l,s)
    \label{matching:sum_k_x}\\
&\quad x^{(s)}_{lk}\in\{0,1\},
\end{align}
\end{subequations}
where $x^{(s)}_{lk}$ being 1 or 0 indicates whether or not $\bpsi_{lk}=\bvphi_s$, the constraint (\ref{matching:sum_t_x}) implies that each user $(l,k)$ can be assigned only one pilot, and the constraint (\ref{matching:sum_k_x}) implies that the users in the same cell cannot be assigned the same pilot.

The weighted bipartite matching problem in (\ref{prob:matching}) is solvable in polynomial time, e.g., by the Hungarian algorithm \cite{hungarian}. After finding the solution of $\underline x$, we recover the solution of the original variables as
\begin{equation}
\label{opt_p_psi}
p^\star_{lk} = \sum^{\tau}_{s=1}x^{(s)}_{lk}p^{(s)}_{lk}
\;\;\text{and}\;\;
\bpsi^\star_{lk} = \sum^{\tau}_{s=1}x^{(s)}_{lk}\bvphi_s.
\end{equation}
The above matching-based optimization is carried out with the auxiliary variable $\underline\bmu$ iteratively updated by (\ref{opt_lambda}).

Because the orthogonal case involves the discrete variable $\underline\bpsi$, it is hard to establish convergence in terms of $\underline\bpsi$. However, the convergence of the objective function can still be guaranteed.

\begin{proposition}
\label{prop:convergence_orthogonal}
The sum of weighted MSEs in (\ref{weighted_MSE}) is monotonically decreasing after each iteration in Algorithm \ref{alg:orthogonal_pilot}.%, and its value converges.
\end{proposition}

Solving the matching problem in (\ref{prob:matching}) incurs cubic computational complexity $O((K+\tau)^3)$. However, this can be simplified to a linear search if we remove the constraint that the users in the same cell cannot be assigned the same pilot, as specified in the following proposition.
\begin{proposition}[Orthogonal Pilot Design via Linear Search]
\label{prop:fast_alg2}
Without the assumption that the users in the same cell cannot be assigned the same pilot, i.e., when constraint (\ref{weighted_MSE_new_orthogonal:d}) is removed, $(\underline p,\underline\bpsi)$ can be optimally determined {\color{black}for the new objective function in (\ref{matching:obj})} as
\begin{equation}
p^\star_{lk} = p^{(s_{lk})}_{lk}\;\;\text{and}\;\;
\bpsi^\star_{lk} = \bvphi_{s_{lk}},
\end{equation}
where the index $s_{lk}$ is obtained by the following linear search:
\begin{equation}
\label{linear_search}
s_{lk} = \arg\max_{s\in[1:\tau]} \pi^{(s)}_{lk}.
\end{equation}
\end{proposition}
The main steps of the proposed orthogonal pilot design are summarized in Algorithm \ref{alg:orthogonal_pilot}.

{\color{black}
\subsection{Max-Min Rate Optimization}
\label{subsec:maxmin}

So far we have focused on weighted MSE minimization. In this section, we extend the problem setting to rate maximization with max-min fairness as in \cite{zhu_wang_dai_qian_15,chien_larsson2018}. Toward this end, we first specialize the rate expression (\ref{rate}) to the orthogonal pilot case in the following corollary which is a well-known result in the literature of massive MIMO.

\begin{corollary}
As $M\rightarrow\infty$, the data rate (\ref{rate}) with orthogonal pilots $p_{lk}\bpsi_{lk}$ reduces to
\begin{equation}
\label{rate2}
{\color{black}R^\infty_{lk}=\log_2\left(1+\frac{\beta_{llk}^2p^2_{lk}}{\sum_{(i,j)\ne(l,k)}\beta_{lij}^2p_{ij}^2\mathds{1}_{\bpsi_{ij}}^{\bpsi_{lk}}}\right),}
\end{equation}
where the indicator variable $\mathds{1}_{\bpsi_{ij}}^{\bpsi_{lk}}$ equals to 1 if $\bpsi_{ij}=\bpsi_{lk}$ and equals to 0 otherwise.
\end{corollary}

Our goal is to maximize the minimum rate $R^\infty_{lk}$ across all users, i.e.,
\begin{subequations}
\label{maxmin:R}
\begin{align}
\underset{\underline{p},\,\underline{\bpsi}}{\text{maximize}} &\quad
  \min_{(l,k)}\big\{R^\infty_{lk}\big\}
    \\
{\text{subject to}}&\quad \text{(\ref{weighted_MSE_new_orthogonal:b})--(\ref{weighted_MSE_new_orthogonal:d})}.
\end{align}
\end{subequations}
Dropping logarithm and substituting
\begin{equation}
\xi_{lk} = p^2_{lk},
\end{equation}
we rewrite (\ref{maxmin:R}) as
\begin{subequations}
\label{prob:xi}
\begin{align}
\underset{\underline{\xi},\,\underline{\bpsi}}{\text{maximize}} &\quad
  \min_{(l,k)}\left\{\frac{\beta_{llk}^2\xi_{lk}}{\sum_{(i,j)\ne(l,k)}\beta_{lij}^2\xi_{ij}\mathds{1}_{\bpsi_{ij}}^{\bpsi_{lk}}}\right\}
    \\
{\text{subject to}}&\quad \text{(\ref{weighted_MSE_new_orthogonal:b})--(\ref{weighted_MSE_new_orthogonal:d})}.
\end{align}
\end{subequations}
We propose to optimize $\underline\xi$ and $\underline \bpsi$ alternatingly. Since the optimization of $\underline \bpsi$ under fixed $\underline \xi$, i.e., orthogonal pilot assignment, has been well studied in the existing literature \cite{zhu_wang_dai_qian_15}, we concentrate on optimizing $\underline\xi$ with $\underline\bpsi$ held fixed. The key step is to recognize (\ref{cor:weighted_MSE_new}) as a concave-convex\footnote{An FP problem is said to be concave-convex if its numerator function is concave while its denominator function is convex.} max-min-ratio problem, so the \emph{generalized Dinkelbach's method} \cite{crouzeix} can be used to find the optimal solution. Specifically, with an auxiliary variable
\begin{equation}
\label{lambda_prime}
\lambda' =  \min_{(l,k)}\left\{\frac{\beta_{llk}^2\xi_{lk}}{\sum_{(i,j)\ne(l,k)}\beta_{lij}^2\xi_{ij}\mathds{1}_{\bpsi_{ij}}^{\bpsi_{lk}}}\right\},
\end{equation}
we decouple the SINRs in (\ref{prob:xi}) as
\begin{subequations}
\label{opt_LP}
\begin{align}
\underset{\underline{\xi}}{\text{maximize}} &\quad
  \min_{(l,k)}\Bigg\{\beta^2_{llk}\xi_{lk}-\sum_{(i,j)\ne(l,k)}\lambda'\beta_{lij}^2\xi_{ij}\mathds{1}_{\bpsi_{ij}}^{\bpsi_{lk}}\Bigg\}
    \\
{\text{subject to}}&\quad 0\le\xi_{lk}\le P^2_{\max}.
\end{align}
\end{subequations}
According to the generalized Dinkelbach's method \cite{crouzeix}, solving the linear programming problem in (\ref{opt_LP}) iteratively leads to the optimal $\underline\xi$ in (\ref{prob:xi}). Algorithm \ref{alg:maxmin} summarizes the above steps.

We remark that the quadratic transform in Theorem \ref{theorem:FP} can be applied to the max-min-ratio (\ref{prob:xi}) as well. The corresponding new problem is
\begin{subequations}
\label{}
\begin{align}
\underset{\underline{\xi}}{\text{maximize}} &\quad
  \min_{(l,k)}\Bigg\{2\lambda_{lk}\beta_{llk}\sqrt{\xi_{lk}}-\sum_{(i,j)\ne(l,k)}\lambda^2_{lk}\beta_{lij}^2\xi_{ij}\mathds{1}_{\bpsi_{ij}}^{\bpsi_{lk}}\Bigg\}
    \\
{\text{subject to}}&\quad 0\le\xi_{lk}\le P^2_{\max},
\end{align}
\end{subequations}
with the auxiliary variable $\lambda_{lk}$ iteratively updated as
\begin{equation}
\label{}
\lambda_{lk}=\frac{\beta_{llk}\sqrt{\xi_{lk}}}{\sum_{(i,j)\ne(l,k)}\beta_{lij}^2\xi_{ij}\mathds{1}_{\bpsi_{ij}}^{\bpsi_{lk}}}.
\end{equation}
Compared to the quadratic transform, the generalized Dinkelbach's method is more efficient here since it introduces only one auxiliary variable $\lambda'$. However, the generalized Dinkelbach's method does not work for a general multi-ratio problem as in (\ref{weighted_MSE_new}).
%We remark that the max-min-ratio problem being concave-convex is critical to the proposed method. For the nonorthogonal pilot case wherein the data rate is computed as (\ref{rate}), the corresponding max-min-ratio problem is no longer concave-convex, so the new problem is still nonconvex and difficult to solve even with each ratio decoupled.

\begin{algorithm}[t]
Initialize $(\underline\bpsi,\underline p)$ to some feasible point\;
\Repeat{the value of $\min_{(l,k)}\gamma_{lk}$ converges}
{Optimize $\underline\bpsi$ via the smart pilot assignment \cite{zhu_wang_dai_qian_15}\;
\Repeat{the value of $\lambda'$ converges}{
    Update the auxiliary variable $\tilde\lambda$ by (\ref{lambda_prime})\;
    Update the power variable $\underline p$ by solving the linear program in (\ref{prob:matching})\;
}}
\caption{Proposed orthogonal pilot design for max-min optimization of data rates}
\label{alg:maxmin}
\end{algorithm}
}

\section{Correlated Rayleigh Fading}
\label{sec:correlated}

This section aims at an extension of the foregoing algorithmic framework to include channel correlation. We now assume that each Rayleigh fading $\bg_{lij}$ is drawn from $\mathcal{CN}(\mathbf0,\bR_{lij})$ where the covariance matrix $\bR_{lij}\in\mathbb C^{M\times M}$ is not necessarily $\bI_M$; other settings remain the same as before.
The MMSE channel estimate now becomes
\begin{equation}
\label{cor:H_est}
\hat\bh_{llk} = \bW_{lk}\bU_l^{-1}\mathrm{vec}\big(\bY_l\big),
\end{equation}
where $\bW_{lk}\in\mathbb C^{M\times\tau M}$ and $\bU_{lk}\in\mathbb C^{\tau M\times\tau M}$ are given by
\begin{equation}
\label{def_W}
\bW_{lk}=\beta_{llk}\bphi^H_{lk}\otimes\bR_{llk}
\end{equation}
and
\begin{equation}
\label{def_U}
\bU_l = \sigma^2\bI_{\tau M}+\sum_{(i,j)}\beta_{lij}\bphi_{ij}\bphi^H_{ij}\otimes\bR_{lij}.
\end{equation}
The resulting MSE is computed as
\begin{equation}
\label{cor:full_MSE}
\mathsf{MSE}_{lk} =\beta_{llk}\mathrm{tr}(\bR_{llk})-\mathrm{tr}\Big(\bW_{lk}\bU_l^{-1}\bW_{lk}^H\Big).
\end{equation}
{\color{black}We remark that similar forms of MSE have been derived in \cite{kotecha_TSP04,liu_wong_hager_TSP07,pang_JSEE08,ottersten_TSP10}, albeit for the single-cell case.}
The correlated version of problem (\ref{weighted_MSE_new}) is therefore
\begin{subequations}
\label{cor:weighted_MSE}
\begin{align}
\underset{\underline\bphi}{\text{maximize}} &\quad
  \sum_{(l,k)}\alpha_{lk}\mathrm{tr}\Big(\bW_{lk}\bU_l^{-1}\bW_{lk}^H\Big).
    \\
{\text{subject to}}&\quad \|\bphi_{lk}\|^2\le \tau P_{\max}.
\end{align}
\end{subequations}
Observe that $\bW_{lk}\bU_l^{-1}\bW_{lk}^H$ is a matrix ratio. In light of the recently developed matrix FP in \cite{matrix_FP}, our ratio-decoupling approach continues to work for (\ref{cor:weighted_MSE_new}), as specified in the following proposition.

\begin{proposition}
The matrix problem in (\ref{cor:weighted_MSE}) is equivalent to
\begin{subequations}
\label{cor:weighted_MSE_new}
\begin{align}
\underset{\underline\bphi,\,\underline\bLambda}{\text{maximize}} &\quad
  f(\underline\bphi,\,\underline\bLambda).
    \\
{\text{subject to}}&\quad \|\bphi_{lk}\|^2\le \tau P_{\max}\\
&\quad \bLambda_{lk}\in\mathbb C^{\tau M\times M},
\end{align}
\end{subequations}
where the new objective function is
\begin{equation}
\label{cor:new_obj}
f(\underline\bphi,\,\underline\bLambda)=\sum_{(l,k)}{\alpha_{lk}}\mathrm{tr}\Big(2\Re\{\bW_{lk}\bm\Lambda_{lk}\}-\bLambda_{lk}^H\bU_l\bLambda_{lk}\Big).
\end{equation}
\end{proposition}
\begin{IEEEproof}
The reformulation is obtained by treating $\bW^H_{lk}$ as $\bA_n(\bx)$ and $\bU_l$ as $\bB_n(\bx)$ in Theorem \ref{theorem:FP_matrix}, along with the nondecreasing function $F_n\big({\mathbf A}^H_n(\bx)\mathbf B_n^{-1}(\bx){\mathbf A}_n(\bx)\big)=\mathrm{tr}\big({\mathbf A}^H_n(\bx)\mathbf B_n^{-1}(\bx){\mathbf A}_n(\bx)\big)$.
\end{IEEEproof}

In an iterative fashion, when $\underline\bphi$ is fixed,
each auxiliary variable $\bLambda_{lk}$ is optimally determined as %follows when $\underline\bphi$ is fixed:
\begin{equation}
\label{def_Lambda}
\bLambda^\star_{lk} =\bU^{-1}_l\bW_{lk}^H.
\end{equation}
This update of $\underline\bLambda$ is optimal regardless of the pilot structure.
Before proceeding to the optimization of $\underline\bphi$ under fixed $\underline\bLambda$, we introduce some shorthand notation:
\begin{itemize}
\item The $m$th row vector of the matrix $\bR_{lij}$ is
\begin{equation}
\bR^m_{lij}=(\be^m_M)^\top\bR_{lij}.
\end{equation}

\item The $s$th $M\times1$ vector on the $m$th column of $\bLambda_{lk}$ is
\begin{equation}
\bLambda^{m,s}_{lk}=\Big(\bE^{[1+(s-1)M:sM]}_{\tau M}\Big)^\top\bLambda_{lk}\be^m_{M}.
\end{equation}

\item The square of $\bLambda_{ij}$ is
\begin{equation}
\widetilde\bLambda_{ij} = \bLambda_{ij}\bLambda^H_{ij}.
\end{equation}

\item The $s$th $M\times1$ vector on the $\big(m+(q+1)M\big)$th column of $\widetilde\bLambda_{ij}$ is
\begin{equation}
\widetilde\bLambda^{m,sq}_{ij} = \Big(\bE^{[1+(s-1)M:sM]}_{\tau M}\Big)^\top\widetilde\bLambda_{ij}\be^{m+(q-1)M}_{M}.
\end{equation}
\end{itemize}

Nonorthogonal pilots and orthogonal pilots are discussed separately in what follows.

\subsubsection{Nonorthogonal Case}

In optimizing nonorthogonal pilots, the central idea is to complete the square for each $\bphi_{lk}$ in the new objective function $f(\underline\bphi,\,\underline\bLambda)$. To this end, we first express  $f(\underline\bphi,\,\underline\bLambda)$ in an alternative form.
\begin{proposition}
\label{prop:correlation_rewrite}
The objective function $f(\underline\bphi,\,\underline\bLambda)$ in (\ref{cor:new_obj}) can be rewritten as
\begin{equation}
\label{cor:f_rewrite}
f(\underline\bphi,\,\underline\bLambda) =
\sum_{(l,k)}2\Re\big\{\bphi_{lk}^H\mathbf v_{lk}\big\}-
\sum_{(l,k)}\bphi_{lk}^H\bQ_{lk}\bphi_{lk}+
\mathrm{const},
\end{equation}
in which $\mathrm{const}$ refers to terms not depending on $\underline\bphi$, the vector variable $\bv_{lk}\in\mathbb C^{\tau}$ is given by
\begin{equation}
\bv_{lk}=\sum^M_{m=1}\alpha_{lk}\beta_{llk}\Big(\bR^m_{llk}\bLambda^{m,1}_{lk},\ldots,\bR^m_{llk}\bLambda^{m,\tau}_{lk}\Big)^\top,
\end{equation}
and the matrix variable $\bQ_{lk}\in\mathbb C^{\tau\times\tau}$ is defined as
\begin{equation}
\label{def_Q}
\bQ_{lk}=\sum\limits_{(i,j,m)}
\alpha_{ij}\beta_{ilk}\left(\begin{matrix}
    \bR^m_{ilk}\widetilde\bLambda^{m,11}_{ij}       & \dots & \bR^m_{ilk}\widetilde\bLambda^{m,1\tau}_{ij} \\
    \vdots       &  & \vdots \\
    \bR^m_{ilk}\widetilde\bLambda^{m,\tau1}_{ij}       & \dots & \bR^m_{ilk}\widetilde\bLambda^{m,\tau\tau}_{ij}
\end{matrix}\right).
\end{equation}
\end{proposition}
\begin{IEEEproof}
See Appendix \ref{appendix:proof}.
\end{IEEEproof}
By completing the square in (\ref{cor:f_rewrite}), the optimal $\bphi_{lk}$ {\color{black}in terms of $f(\underline\bphi,\,\underline\bLambda)$} can be readily obtained as
\begin{equation}
\label{cor:opt_phi}
\bphi^\star_{lk} = \big(\bQ_{lk}+\eta_{lk}\bm I_{\tau M}\big)^{-1}\bv_{lk},
\end{equation}
where the Lagrange multiplier $\eta_{lk}$ is again determined by (\ref{opt_eta}). Furthermore, we can make use of Proposition \ref{prop:fast_alg1} to simplify the update of $\bphi_{lk}$: when the SNR is sufficiently high, we just scale the pilots properly to meet the power constraint, thus getting rid of the Lagrange multiplier $\eta_{lk}$.

The convergence of Algorithm \ref{alg:arbitrary_pilot} as stated in Proposition \ref{prop:convergence} carries over to this correlated channel case.
%This alterative method of optimizing $\bphi_{lk}$ is far more computationally efficient since it avoids a sequence of matrix inversion in (\ref{cor:opt_phi})

% the scaling method in Proposition \ref{prop:fast_alg1} can be carried over to the current case directly so that we do not need to compute $\eta_{lk}$ when optimizing $\bphi^\star_{lk}$. Like the uncorrelated channel case, this scaling method is equivalent to (\ref{cor:opt_phi}) when the noise level $\sigma^2$ tends to zero.

\begin{table*}[t]
\renewcommand{\arraystretch}{1.3}
\small
\centering
{
\caption{\small Computational Complexity and Communication Complexity of Proposed Algorithms}
{\color{black}
\begin{tabular}{|c||c|c|c|c|}
\hline
&\multicolumn{2}{c|}{Uncorrelated Channel Case}&\multicolumn{2}{c|}{Correlated Channel Case}\\
\hline
& Computational Complexity & Communication Complexity & Computational Complexity & Communication Complexity \\
\hline
\hline
Algorithm \ref{alg:arbitrary_pilot} & $O(K^2L^2\tau^2+KL\tau^3)$ & $O(KL^2\tau)$ & $O(K^2L^2M^2\tau^2+KL\tau^3)$ & $O(KL^2M^2\tau)$  \\
\hline
Algorithm \ref{alg:orthogonal_pilot} & $O(K^2L^2\tau^3+K^3L)$ &  $O(KL^2\tau)$ & $O(K^2L^2M^2\tau^3+K^3L)$ & $O(KL^2M^2\tau)$\\
\hline
Algorithm \ref{alg:maxmin} & $O(K^{2.5}L^{2.5})$ &  $O(KL^2\log(\tau))$ & -- &  --\\
\hline
\end{tabular}
\label{tab:complexity}
}}
\end{table*}

\subsubsection{Orthogonal Case}

We next generalize the orthogonal pilot design to correlated Rayleigh fading. The main procedure here follows that of Section \ref{sec:orthogonal}. Replacing $\bphi_{lk}$ with $(p_{lk},\bpsi_{lk})$ in (\ref{cor:f_rewrite}), we express the objective function of the orthogonal pilots in (\ref{orthogonal_pilot}) as
\begin{multline}
\label{cor:f_rewrite_orthogonal}
f(\underline p,\,\underline\bpsi\,,\underline\bLambda) =
\sum_{(l,k)}2\sqrt{p_{lk}}\,\Re\big\{\bpsi_{lk}^H\mathbf v_{lk}\big\}\\
-\sum_{(l,k)}p_{lk}\bpsi_{lk}^H\bQ_{lk}\bpsi_{lk}+
\mathrm{const},
\end{multline}
where $\mathrm{const}$ refers to terms not depending on $(\underline p,\underline\bpsi)$, and $\bQ_{lk}$ is defined in (\ref{def_Q}).

If a particular normalized pilot $\bvphi_s$ is assigned to user $(l,k)$, the corresponding optimal $p_{lk}$ {\color{black}for $f(\underline p,\,\underline\bpsi\,,\underline\bLambda)$} is given by
\begin{equation}
p^{(s)}_{lk} = \frac{\Re\big\{\bvphi^H_{s}\bv_{lk}\big\}}{\bvphi^H_{s}\bQ_{lk}\bvphi_{s}}.
\end{equation}
The contribution of user $(l,k)$ to $f(\underline p,\,\underline\bpsi\,,\underline\bLambda)$  is then computed as
\begin{equation}
\label{new_weight}
\pi^{(s)}_{lk} =
2\sqrt{p^{(s)}_{lk}}\Re\big\{\bvphi^H_{s}\bv_{lk}\big\}\,-\\
p^{(s)}_{lk}\bvphi^H_{s}\bQ_{lk}\bvphi_{s}.
\end{equation}
We aim to find the optimal assignment of $\{\bvphi_1,\ldots,\bvphi_{\tau}\}$  such that value of $f(\underline p,\,\underline\bpsi\,,\underline\bLambda)$ is maximized. This target can be reached by solving the same weighted bipartite matching problem as in (\ref{prob:matching}) except that the link weight is evaluated as (\ref{new_weight}). Again, if we allow the users in the same cell to be assigned the same pilot, each user $(l,k)$ simply chooses its $(p_{lk},\bpsi_{lk})$ according to $\pi^{(s)}_{lk}$ by linear search. The property of Algorithm \ref{alg:orthogonal_pilot} stated in Proposition \ref{prop:convergence_orthogonal} continues to hold in the correlated channel case.

{\color{black}In contrast, the max-min-rate method in Algorithm \ref{alg:maxmin} cannot be extended for the correlated channel case because the max-min-ratio problem in (\ref{prob:xi}) is no longer concave-convex.}

{\color{black}
\section{Complexity Analysis}
\label{sec:complexity}

This section examines how the computational complexities and communication complexities of the proposed algorithms scale with the system parameters $(K,L,M,\tau)$. Our discussion focuses on the general form of each algorithm, e.g., Option 1 of Algorithm \ref{alg:arbitrary_pilot}.

We begin with computational complexities. The following analysis is for each iteration in the proposed algorithms:
\begin{itemize}
\item\emph{Algorithm \ref{alg:arbitrary_pilot}:} First, it requires a computational complexity of $O(KL\tau^2)$ to compute each $\bD_l$ in (\ref{def_D}), then $O(\tau^3)$ operations to compute the inverse matrix $\bD_l^{-1}$ in (\ref{opt_lambda}); because the above operations are done for each cell $l$, the overall complexity is $O(KL^2\tau^2+L\tau^3)$. Second, it requires $O(\tau^2)$ operations to compute each $\bm\lambda_{lk}$ in (\ref{opt_lambda}), so that the overall complexity across a total of $KL$ users is $O(KL\tau^2)$. Third, it requires $O(KL\tau^2+\tau^3)$ operations to compute each $\bphi_{lk}$ in (\ref{opt_phi}), leading to the overall computational complexity across all users of $O(K^2L^2\tau^2+KL\tau^3)$.
    Summarizing, the overall computational complexity is $O(KL^2\tau^2+KL\tau^3)$. Likewise, it can be shown that Algorithm \ref{alg:arbitrary_pilot} for the correlated channel case has a computational complexity of $O(KL^2M^2\tau^2+KLM^3\tau^3)$.

\item\emph{Algorithm \ref{alg:orthogonal_pilot}:} The update of $\underline{\bm\lambda}$ in Algorithm \ref{alg:orthogonal_pilot} is the same as that in Algorithm \ref{alg:arbitrary_pilot}, with a computational complexity of $O(KL^2\tau^2+L\tau^3)$. We focus on the matching part. It requires $O(KL\tau^2)$ operations to compute the weight $\pi^{(s)}_{lk}$ with respect to each $(l,k,s)$, leading to a total computational complexity $O(K^2L^2\tau^3)$. The subsequent weighted bipartite matching requires $O((\tau+K)^3)$ operations per cell. Thus, the overall computational complexity of Algorithm \ref{alg:orthogonal_pilot} is $O(K^2L^2\tau^3)+O((\tau+K)^3L)=O(K^2L^2\tau^3+K^3L)$. When extended to the correlated channel case, the algorithm requires a computational complexity of $O(K^2L^2M^2\tau^3+K^3L)$.

\item\emph{Algorithm \ref{alg:maxmin}:} First, it requires $O(K^2L^2)$ operations to compute the auxiliary variable $\lambda'$ in (\ref{lambda_prime}). We then update $\underline p$ by solving a linear programming problem with $KL$ scalar variables. According to the classic work \cite{vaidya_focs}, the problem can be solved with a computational complexity of $O(K^{2.5}L^{2.5})$.
\end{itemize}

Next we consider the communication complexities. It is assumed that every BS $l$ knows \emph{a priori} the large-scale fading related to its cell, i.e., $\{\beta_{lij},\forall (i,j)\}$ and $\{\beta_{ilk},\forall (i,k)\}$, and the correlation matrices $\{\bR_{lij},\forall (i,j)\}$ and $\{\bR_{ilk},\forall (i,k)\}$ in addition for the correlated channel case.
Our analysis focuses on the communication of the pilot variable, the power variable, and the auxiliary variable between different cells.

\begin{itemize}
\item\emph{Algorithm \ref{alg:arbitrary_pilot}:} Each BS $l$ needs to let other BSs know its pilot variable $\{\bphi_{lk}\in\mathbb C^\tau,k\in[1:K]\}$ and auxiliary variable $\{\bm\lambda_{lk}\in\mathbb C^\tau,k\in[1:K]\}$; the resulting total communication complexity across $L$ cells is $O(KL^2\tau)$. In the correlated channel case, the auxiliary variable becomes $\{\bm\Lambda_{lk}\in\mathbb C^{\tau M\times M},k\in[1:K]\}$, so that the communication complexity rises to $O(KL^2M^2\tau)$.
\item\emph{Algorithm \ref{alg:orthogonal_pilot}:} Each BS $l$ needs to let other BSs know the orthogonal pilot indices $s\in[1:\tau]$ used in its cell, leading to a total communication complexity across $L$ cells of $O(KL^2\log(\tau))$. Likewise, the total cost of communicating the power variable $\underline p$ is $O(KL^2)$. The communication of the auxiliary variable $\underline{\bm\lambda}$ has the same complexity as in Algorithm \ref{alg:arbitrary_pilot}. Thus, the overall communication complexity of Algorithm \ref{alg:orthogonal_pilot} is still $O(KL^2\tau)$. The correlated channel case requires $O(KL^2M^2\tau)$.
\item\emph{Algorithm \ref{alg:maxmin}:} Differing from Algorithm \ref{alg:arbitrary_pilot} or \ref{alg:orthogonal_pilot} that lets each cell update its own variables in a distributed fashion, Algorithm \ref{alg:maxmin} employs a central controller to optimize the entire network. Because the auxiliary variable $\lambda'$ is computed locally at the central controller, the communication cost is only caused by $\underline{\bpsi}$ and $\underline p$, which amounts to $O(KL\log(\tau))$.
\end{itemize}

The above complexity results are summarized in Table \ref{tab:complexity}.
}

\section{Numerical Results}
\label{sec:numerical}

We validate the performance of the proposed algorithms in a wireless network with 7 hexagon-shape cells wrapped around. In each cell, there is one BS located at the center and 6 users uniformly distributed. The  BS-to-BS distance equals 500m. We assume that each BS has 100 antennas and that a $20$MHz-wide spectrum band is reused in each cell. {\color{black}The channel model follows ITU-R M.1225 PedA \cite{wei_infocom}:} the pathloss is computed as $128.1+37.6\log_{10}(d)$ in dB, where $d$ refers to the distance in km; the shadowing in dB between any pair of transmitter and receiver is modeled as an i.i.d. Gaussian random variable drawn from {\color{black}$\mathcal{N}(0,64)$}. Moreover, we set the background noise $\sigma^2$ at $-169$dBm/Hz \cite{bjornson_15}, and set the maximum transmit power level $P_{\max}$ at $23$dBm \cite{LTE_book}.

%The other parameters follow: $M=100$, $K=6$, $\tau=10$, $\sigma^2=-100$dBm, and $P_{\max}=23$dBm. The parameter setting will be changed slightly when we move on to the correlated fading case, as stated prior to Fig.~\ref{fig:correlated_iter} and Fig.~\ref{fig:correlated_distr}.

{\color{black}
In the rest of this section, Algorithm \ref{alg:arbitrary_pilot}, Algorithm \ref{alg:orthogonal_pilot}, and Algorithm \ref{alg:maxmin} are referred to as nonorthogonal FP, orthogonal FP, and max-min FP, respectively, and are compared with the following benchmarks:
\begin{itemize}
\item\emph{Orthogonal Method:} Fix a set of $\tau$ orthogonal pilots, and randomly assign a subset of $K$ orthogonal pilots to each cell; the pilots are all transmitted at full power $P_{\max}$.
\item\emph{Random Method:} Generate each pilot symbol randomly and independently according to the Gaussian distribution under the maximum power constraint.
\item\emph{GSRTM \cite{eldar_18}:} This method optimizes the nonorthogonal pilot symbols successively in order to minimize the sum MSEs of channel estimation.
\item\emph{Lower Bound:} We obtain a lower bound on the sum MSEs of channel estimation by ignoring pilot contamination.
\item\emph{Smart Pilot Assignment \cite{zhu_wang_dai_qian_15}:} It assigns the set of fixed orthogonal pilots to the users in each cell in a greedy fashion iteratively.
\end{itemize}
}
{\color{black}We do not compare with the method in \cite{chien_larsson2018}, because the complexity involved in successive geometric programming only allows very short pilots to be designed.}
%For ease of implementation, we assume by default that \emph{Option 2} is adopted in both Algorithm \ref{alg:arbitrary_pilot} (i.e., without the Lagrangian multiplier) and Algorithm \ref{alg:orthogonal_pilot} (i.e., based on linear search). In addition to the GSRTM method based on a random dictionary \cite{eldar_18}, the successive approximation method \cite{salihi_nakhai_2018}, and the smart orthogonal pilot assignment \cite{zhu_wang_dai_qian_15}, we include two more benchmark methods:
%\begin{itemize}
%\item\emph{Orthogonal Method:} Fix a set of 10 orthogonal pilots at the max power; select 6 pilots randomly in each cell;
%\item\emph{Random Method:} Generate the pilots randomly and independently according to the Gaussian distribution.
%\item\emph{GSRTM:}
%\item\emph{Lower Bound:}
%\end{itemize}
%\vspace{-0.35em}
%The orthogonal method is used to initialize the other methods if a starting point is needed. We remark that GSRTM \cite{eldar_18}, successive approximation \cite{salihi_nakhai_2018}, the random method, and Algorithm \ref{alg:arbitrary_pilot} aim at the arbitrary pilot design, while the rest aim at the orthogonal pilot design.

% For data signals, we adopt the channel inversion power control in \cite{bjornson_larsson_debbah_twc16}, i.e., $\tilde{\rho}_{ik}=\delta/\beta_{i,ik}$ and we set $\delta=1$ without loss of generality.
\begin{figure}[t]
\begin{minipage}[b]{1.0\linewidth}
  \centering
  \centerline{\includegraphics[width=9.5cm]{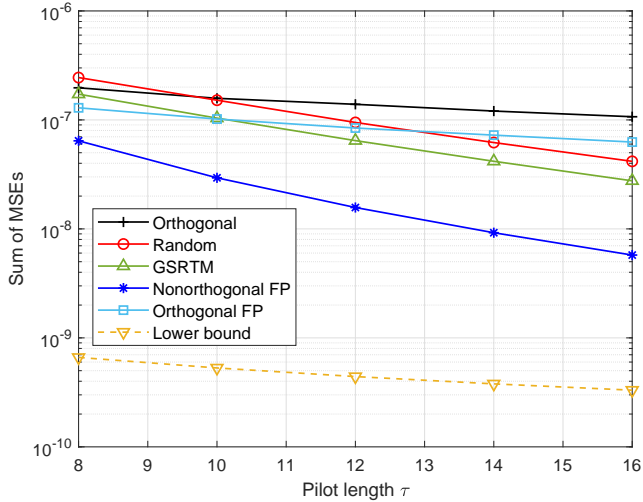}}
  \caption{Sum MSEs minimization in the uncorrelated channel case.}
  \label{fig:sum_MSE}
\end{minipage}
\end{figure}

\begin{figure}[t]
\begin{minipage}[b]{1.0\linewidth}
  \centering
  \centerline{\includegraphics[width=9.5cm]{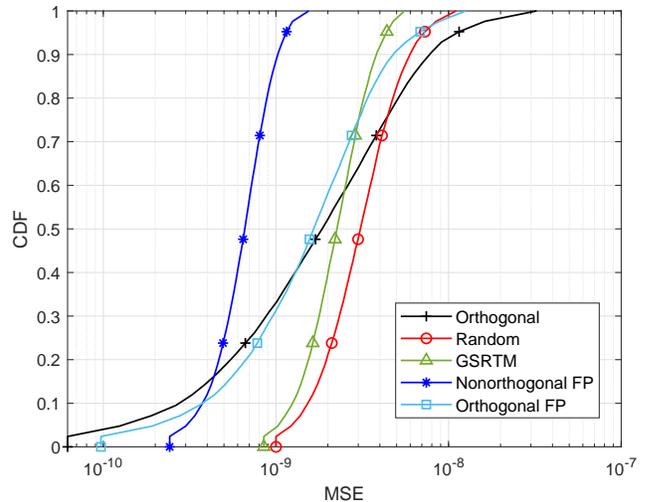}}
  \caption{CDF of MSEs in the uncorrelated channel case.}
  \label{fig:cdf_MSE}
\end{minipage}
\end{figure}

\begin{figure}[t]
\begin{minipage}[b]{1.0\linewidth}
  \centering
  \centerline{\includegraphics[width=9.5cm]{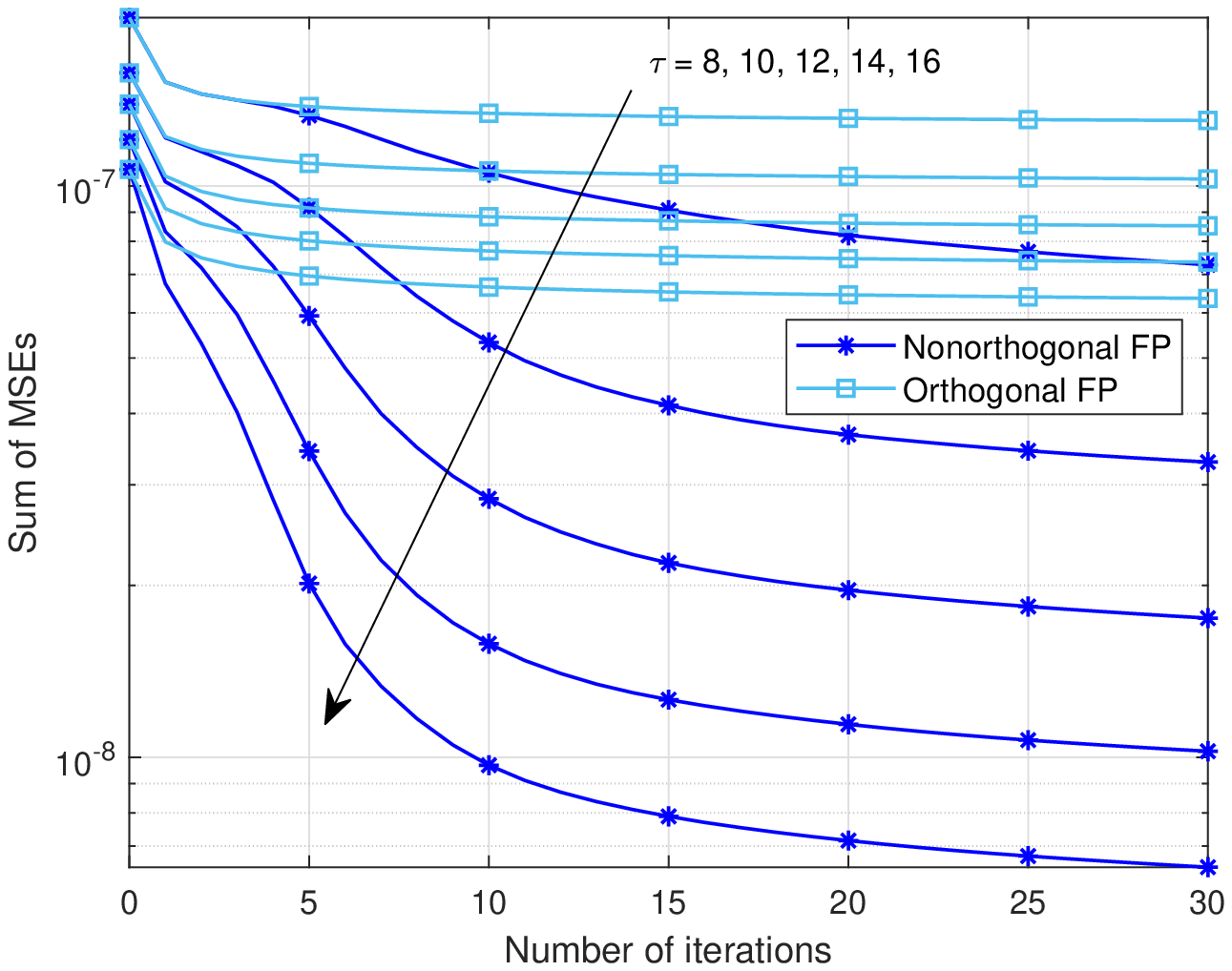}}
  \caption{Convergence of sum MSEs in the uncorrelated channel case.}
  \label{fig:convergence_MSE}
\end{minipage}
\end{figure}

\begin{figure}[t]
\begin{minipage}[b]{1.0\linewidth}
  \centering
  \centerline{\includegraphics[width=9.5cm]{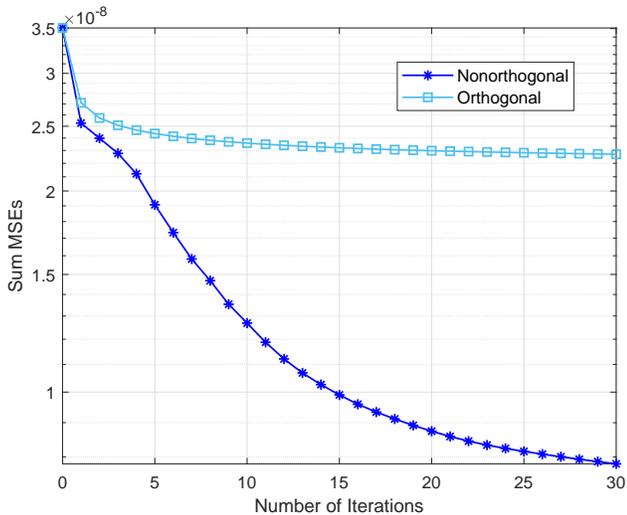}}
  \caption{Convergence of sum MSEs in the correlated channel case.}
  \label{fig:cor_MSE}
\end{minipage}
\end{figure}

We first consider minimizing the sum MSEs for uncorrelated channels, i.e., problem (\ref{weighted_MSE_new}) with each $\alpha_{lk}$ set to 1. As shown in Fig.~\ref{fig:sum_MSE}, the proposed nonorthogonal FP method outperforms the other techniques significantly, e.g., as compared to the random method, it reduces the sum MSEs by around 10dB when $\tau=16$. The figure shows that the algorithms using nonorthogonal pilots tend to yield lower sum MSEs. But this is not always the case as we observe that the GSRTM method and the random method are inferior to the proposed orthogonal FP approach when $\tau$ is sufficiently small. Observe also that the random technique is even worse than the orthogonal method at $\tau=8$. Thus, nonorthogonal pilot design becomes quite crucial when short pilots are used.

{\color{black}Fig.~\ref{fig:cdf_MSE} further shows the cumulative distribution function (CDF) of sum MSEs with respect to a large number of random network realizations. Remarkably, the nonorthogonal FP method achieves smaller sum MSEs than the random and GSRTM techniques at any percentile. The figure also shows that the two orthogonal approaches, the orthogonal FP method and the orthogonal baseline, outperform the nonorthogonal FP algorithm only in the low MSE regime, but are much worse elsewhere.}

Fig.~\ref{fig:convergence_MSE} shows the convergence of sum MSEs for the proposed FP methods. The two algorithms are both initialized by the orthogonal baseline. Observe that the orthogonal FP method converges after only 10 iterations. In comparison, it takes many more iterations for the nonorthogonal FP approach to converge, but the improvement achieved in the first few iterations is considerable, e.g., the sum MSEs is already reduced by 10dB after 10 iterations $\tau=16$.

\begin{figure}[t]
\begin{minipage}[b]{1.0\linewidth}
  \centering
  \centerline{\includegraphics[width=9.5cm]{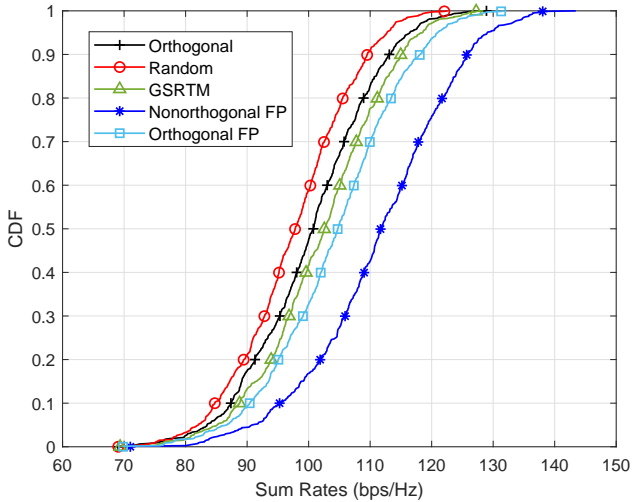}}
  \caption{CDF of sum rates in the uncorrelated channel case.}
  \label{fig:rate_MSE}
\end{minipage}
\end{figure}

We further plot in Fig.~\ref{fig:cor_MSE} the convergence of sum MSEs for a correlated channel case with $\tau=5$. For ease of simulations, we set $K=3$, the other parameters remaining the same as before. The channel covariance matrix $\bR_{lij}$ is obtained from the exponential model in \cite{loyka,chien_larsson2018}: randomly generate $\omega_{lij}=\nu e^{\mathsf j \theta}$ wherein $\nu$ is set to $0.5$ and $\theta$ is drawn i.i.d. from the uniform distribution $U[0,2\pi)$, then set the $(m,n)$th entry of the matrix $\bR_{lij}$ as
\begin{equation}
\label{generate_R}
	\bR^{m,n}_{lij} =
\left\{
\begin{aligned}
& \omega^{m-n}_{lij},\;\text{if $m\ge n$};\\
&(\bR^{m,n}_{lij})^H,\; \text{otherwise.}
\end{aligned}
\right.
\end{equation}
The convergence shown in Fig.~\ref{fig:cor_MSE} is slightly slower than that of the uncorrelated channel case, but their overall profiles are similar.

{\color{black}
The remainder of this section considers uncorrelated channels along with $K=6$ and $\tau=10$. We now compare the sum rate performance of the different methods, assuming that pilots have been optimized for the normalized sum MSEs with $\alpha_{lk}=1/\beta_{llk}$ \cite{liu_jiang_TCOM17,Hanzo_STSP14,Shariati_STSP14,bogale_le_15}. According to Fig.~\ref{fig:rate_MSE}, the nonorthogonal FP algorithm improves upon the other techniques by about 10\% at the 50th percentile, even without using any sophisticated power control and receiver design as assumed in (\ref{rate}). It is also worthwhile to note that the orthogonal FP approach is superior to GSRTM, although the latter allows arbitrary nonorthogonal pilots.

Finally, we evaluate the performance of the max-min FP approach. For simplicity, the outer-loop iteration in Algorithm \ref{alg:maxmin} is run only once. Despite this simplification, the max-min FP method can already significantly outperform the smart pilot assignment in \cite{zhu_wang_dai_qian_15}. For example, max-min FP reaches a minimum rate higher than 2 Mbps/Hz with probability 58\%, whereas the smart pilot assignment only with 10\%.
}
%By contrast, Algorithm \ref{alg:arbitrary_pilot} with normalizing weights has a fairly different profile. It puts main efforts in reducing MSEs for the weak users, albeit at the cost of those strong users which can tolerate larger estimation error. Thus, using the normalized MSE as criterion is more reasonable in practice.

\begin{figure}[t]
\begin{minipage}[b]{1.0\linewidth}
  \centering
  \centerline{\includegraphics[width=9.5cm]{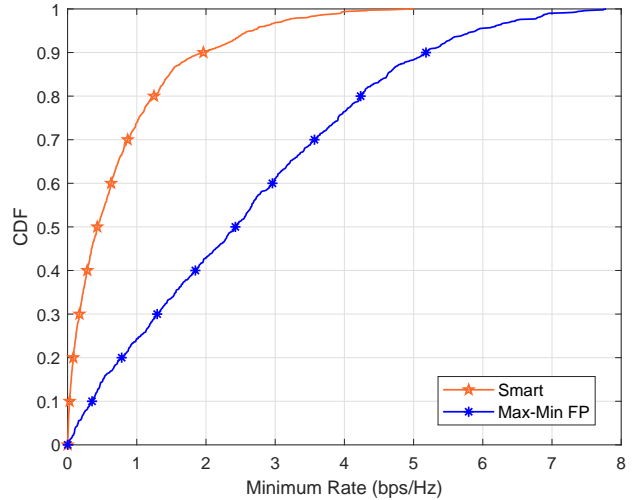}}
  \caption{Max-min-rate optimization in the uncorrelated channel case.}
  \label{fig:maxmin_rate}
\end{minipage}
\end{figure}

\section{Conclusion}
\label{sec:conclusion}

This work proposes an FP framework for coordinating the uplink pilots across multiple cells in order to mitigate pilot contamination in massive MIMO. This approach produces a closed-form method for the nonorthogonal pilot design, and a weighted bipartite matching for orthogonal pilot assignment and power control. Further extensions to the max-min-rate optimization and the correlated channel estimation are obtained using max-min-ratio FP and matrix FP, respectively. Numerical results show that the proposed algorithms are superior to the state-of-the-art techniques for both weighted MSE minimization and max-min-rate optimization.

\appendices

{\color{black}
\section{Proof of Theorem \ref{theorem:rate}}
\label{appendix:A}

Let $\tilde x_{lk}\sim\mathcal{CN}(0,1)$ be the data signal transmitted from user $(l,k)$ and let $\tilde\bz\sim\mathcal{CN}(\mathbf0,\sigma\bI_M)$ be the AWGN at BS $l$ during the uplink transmission. With respect to each user $(l,k)$, BS $l$ uses the conjugate transpose of the corresponding channel estimation $\hat\bh_{llk}$ as the maximum-ratio combining (MRC) receiver, thus obtaining
\begin{align}
\label{tilde_y1}
\tilde{y}_{lk}&= \hat\bh_{llk}^H\Bigg(\bh_{llk}\tilde{x}_{lk}+\sum_{(i,j)\ne(l,k)}\bh_{lij}\tilde x_{ij}+\tilde\bz_{l}\Bigg).
\end{align}
Because the exact value of $\bh_{llk}$ is unknown at BS $l$, it is difficult to compute the achievable rate based on (\ref{tilde_y1}) directly. Following the use-and-then-forget method in \cite{massiveMIMObook2016}, we introduce the correlation variable
\begin{equation}
\mu_{lk} = \mathbb E\big[\hat\bh_{llk}^H\bh_{llk}\big],
\end{equation}
and rewrite (\ref{tilde_y1}) as
\begin{align}
\label{tilde_y}
\tilde{y}_{lk}&= \mu_{lk}\tilde x_{lk}+\Delta_{lk},
\end{align}
where
\begin{equation}
\Delta_{lk} =\sum_{(i,j)}\hat{\bh}^H_{lij}\bh_{lij}\tilde{x}_{ij}+\hat{\bh}^H_{llk}\tilde{\bz}_{lk}-\mu_{lk}\tilde{x}_{lk}.
\end{equation}
It turns out that $\bbE[(\mu_{lk}x_{lk})^H\Delta_{ik}]=0$ (with the expectation taken over all random variables from channel estimation and data transmission, i.e., $\underline\bh,\underline\bZ,\underline{\tilde x},\underline{\tilde\bz}$), so (\ref{tilde_y}) can be treated as if $\tilde x_{lk}$ passed through the channel $\mu$ with added uncorrelated noise $\Delta_{lk}$.

Most importantly, the exact value of $\mu_{lk}$ is available to BS $l$ so long as it knows how the channels and noise are distributed. Henceforth, according to (\ref{tilde_y}), the following data rate is achievable for user $(i,k)$:
\begin{align}
\label{lower_bound}
{R}_{lk} &=\log_2\Bigg(1+|\mu_{lk}|^2\cdot\frac{\bbE\big[|\tilde x_{lk}|^2\big]}{\bbE\big[|\Delta_{lk}|^2\big]}\Bigg),
\end{align}
which can be rewritten as in (\ref{rate}) after some algebra.
}

\section{Proof of Proposition \ref{prop:correlation_rewrite}}
\label{appendix:proof}

We first introduce a lemma used to simplify the calculation with a Kronecker product.

\begin{lemma}
\label{lemma:kronecker}
The following identity holds true given any $\mathbf a\in\mathbb C^{n_1}$, $\mathbf b\in\mathbb C^{n_2}$, $\mathbf C\in\mathbb C^{n_3\times n_4}$, and $\mathbf F\in\mathbb C^{n_2n_4\times n_1n_3}$:
\begin{equation}
\mathrm{tr}\Big(\big((\mathbf a\mathbf b^H)\otimes\mathbf C\big)\mathbf F\Big) = \mathbf b^H\mathbf T\,\mathbf a,
\end{equation}
where the $(i,j)$th entry of $\mathbf T\in\mathbb C^{n_2\times n_1}$ is computed as
\begin{equation}
T_{ij} = \sum^{n_3}_{m=1}(\be^m_{n_3})^\top\mathbf C\big(\bE^{[1+(i-1)n_4:in_4]}_{n_3}\big)^\top\mathbf F\,\be^{j+(m-1)n_1}_{n_1n_3}.
\end{equation}
Observe that $(\be^m_{n_3})^\top\mathbf C$ corresponds to the $m$th row of $\mathbf C$ while $\big(\bE^{[1+(i-1)n_4:in_4]}_{n_3}\big)^\top\mathbf F\,\be^{j+(m-1)n_1}_{n_1n_3}$ corresponds to the $i$th $n_4\times1$ vector on the $\big(j+(m-1)n_1\big)$th column of $\mathbf F$. The proof is based on expanding the Kronecker product $(\mathbf a\mathbf b^H)\otimes\mathbf C$, followed by some elementary linear algebra.
\end{lemma}

We now return to the new objective function $f(\underline\bphi,\,\underline\bLambda)$ in (\ref{cor:new_obj}). Its positive terms can be rewritten as
\begin{subequations}
\begin{align}
&\sum_{(l,k)}{\alpha_{lk}}\mathrm{tr}\Big(2\Re\{\bW_{lk}\bm\Lambda_{lk}\}\Big)\notag\\
&=\sum_{(l,k)}{\alpha_{lk}}\mathrm{tr}\Big(2\Re\{\beta_{llk}\bphi^H_{lk}\otimes\bR_{llk}\bm\Lambda_{lk}\}\Big)\\
&=\sum_{(l,k)}2\Re\Big\{\mathrm{tr}\Big(\big({\alpha_{lk}}\beta_{llk}\bphi^H_{lk}\otimes\bR_{llk}\big)\bm\Lambda_{lk}\Big)\Big\}\\
&=\sum_{(l,k)}2\Re\big\{\bphi_{lk}^H\bv_{lk}\big\},
    \label{rewrite_pos}
\end{align}
\end{subequations}
where the last equality is due to Lemma \ref{lemma:kronecker} with $\mathbf a$, $\mathbf b$, $\mathbf C$, and $\mathbf F$ set to $2\alpha_{lk}\beta_{llk}\bphi_{lk}^H$, $1$, $\bR_{llk}$, and $\bLambda_{lk}$, respectively.
Furthermore, the negative terms of $f(\underline\bphi,\,\underline\bLambda)$ can be rewritten as
\allowdisplaybreaks
\begin{subequations}
\begin{align}
&\sum_{(l,k)}{\alpha_{lk}}\mathrm{tr}\Big(\bLambda_{lk}^H\bU_l\bLambda_{lk}\Big)\notag\\
&=\sum_{(l,k)}{\alpha_{lk}}\mathrm{tr}\Big(\bU_l\widetilde{\bLambda}_{lk}\Big)\\
&=\sum_{(l,k)}{\alpha_{lk}}\mathrm{tr}\Bigg(\sum_{(i,j)}\Big(\big(\beta_{lij}\bphi_{ij}\bphi^H_{ij}\big)\otimes\bR_{lij}\Big)\widetilde{\bLambda}_{lk}\Bigg)+\mathrm{const}\\
&\overset{(*)}{=}\sum_{(l,k)}{\alpha_{lk}}\Bigg(\sum_{(i,j)}\beta_{lij}\Big(\bphi_{ij}^H\mathbf T_{lij}\bphi_{ij}\Big)\Bigg)+\mathrm{const}\\
&=\sum_{(l,k)}\bphi_{lk}^H\mathrm{tr}\Bigg(\sum_{(i,j)}\alpha_{ij}\beta_{ilk}\mathbf T_{ilk}\Bigg)\bphi_{lk}+\mathrm{const}\\
&=\sum_{(l,k)}\bphi_{lk}^H\bQ_{lk}\bphi_{lk}+\mathrm{const},
    \label{rewrite_neg}
\end{align}
\end{subequations}
where $\mathrm{const}=\tau \sigma^2\sum_{(l,k)}\alpha_{lk}\mathrm{tr}(\widetilde\bLambda_{lk})$ does not depend on $\underline\bphi$; step $(*)$ follows Lemma \ref{lemma:kronecker} by letting $\mathbf a=\mathbf b=\bphi_{ij}$, $\mathbf C=\bR_{lij}$, and $\mathbf F=\widetilde\bLambda_{lk}$. Combining (\ref{rewrite_pos}) and (\ref{rewrite_neg}) gives the new form of $f(\underline\bphi,\,\underline\bLambda)$ in (\ref{cor:f_rewrite}).

\IEEEpeerreviewmaketitle

\bibliographystyle{IEEEbib}
\bibliography{IEEEabrv,strings}

\begin{IEEEbiography}
[{\includegraphics[width=1in,height=1.25in,clip,keepaspectratio]{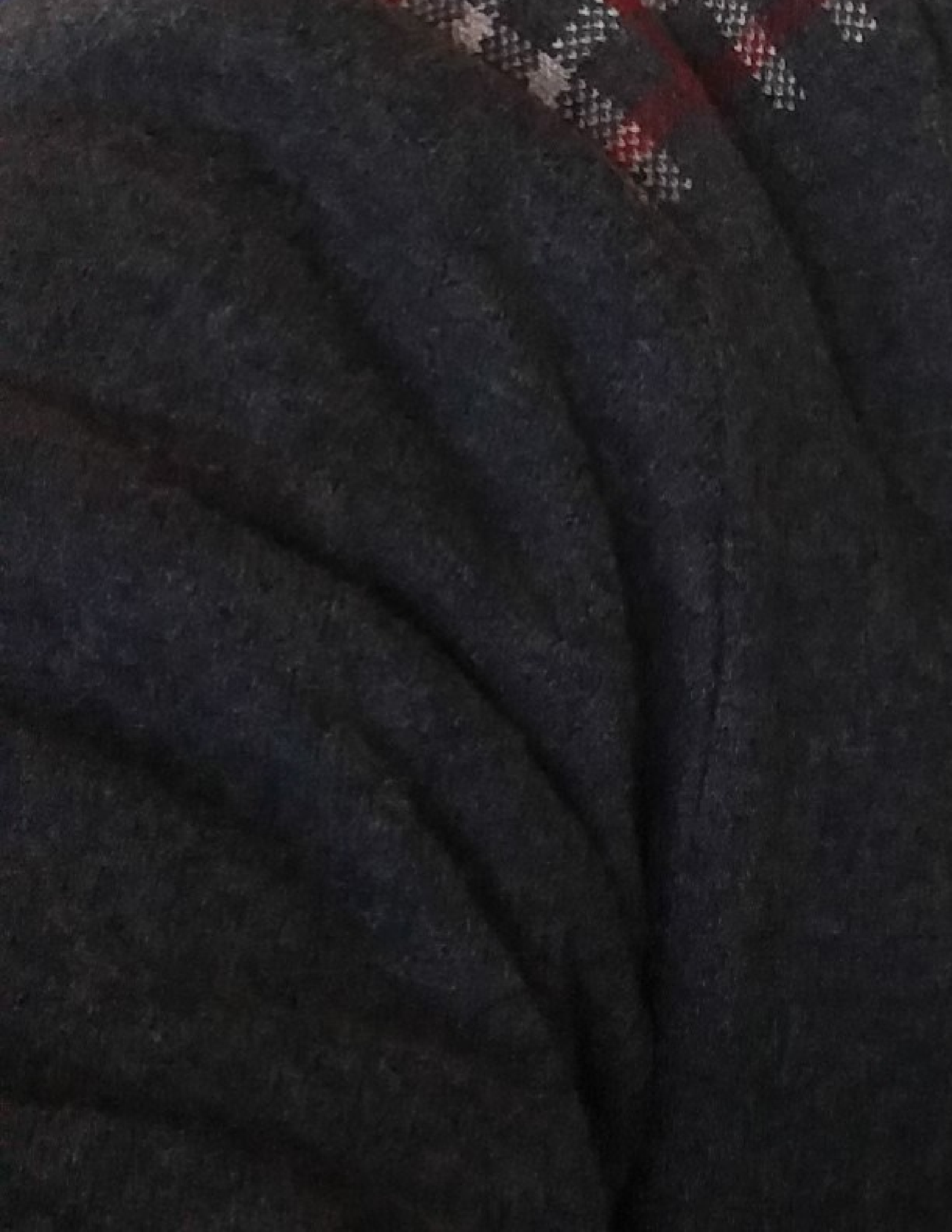}}]
{Kaiming Shen}
(S'13) received the B.Eng. degree in information security and the B.S. degree in mathematics from Shanghai Jiao Tong University, Shanghai, China in 2011, then the M.A.Sc. and Ph.D. degrees in electrical and computer engineering from the University of Toronto, Ontario, Canada in 2013 and 2020, respectively.

Since 2020, he has been an Assistant Professor with the School of Science and Engineering at the Chinese University of Hong Kong (Shenzhen), Shenzhen, Guangdong, China. His main research interests include optimization, wireless communications, data science, and information theory.
\end{IEEEbiography}

\begin{IEEEbiography}
[{\includegraphics[width=1in,height=1.25in,clip,keepaspectratio]{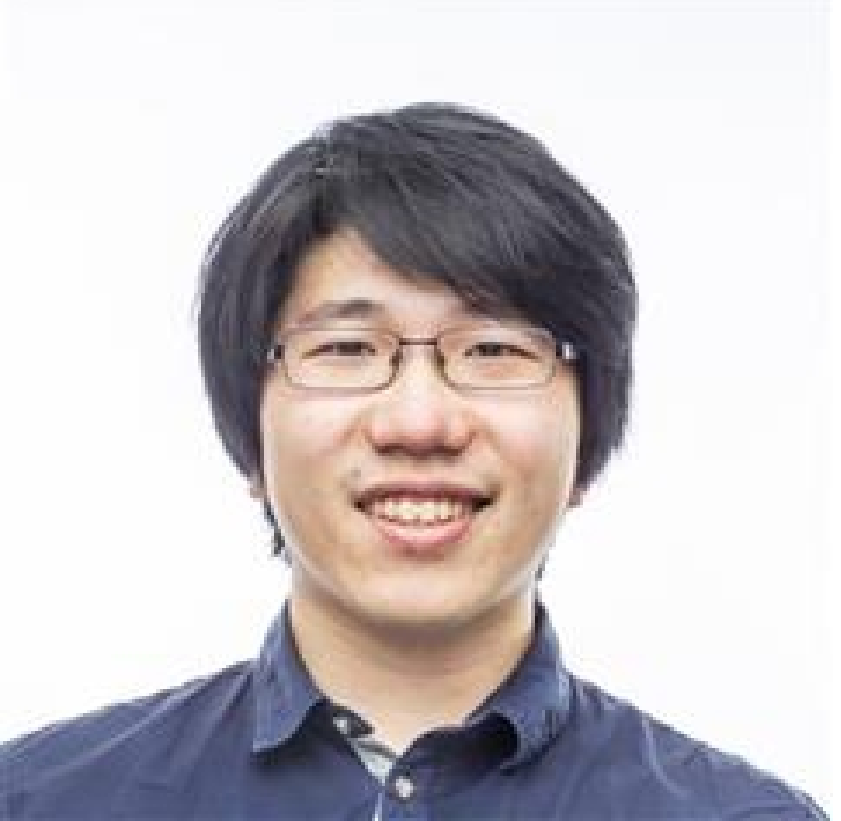}}]
{Hei Victor Cheng}
(S'13-M'18) received the B.Eng. degree in electronic engineering from Tsinghua University, Beijing, China, the M.Phil. degree in electronic and computer engineering from The Hong Kong University of Science and Technology, and the Ph.D. degree from the Department of Electrical Engineering, Linköping University, Sweden. He is currently a Post-Doctoral Research Fellow with the University of Toronto, Canada. His current research interests include Massive MIMO, statistical signal processing, optimization theory, and machine learning for communications.
\end{IEEEbiography}

\begin{IEEEbiography}
[{\includegraphics[width=1in,height=1.25in,clip,keepaspectratio]{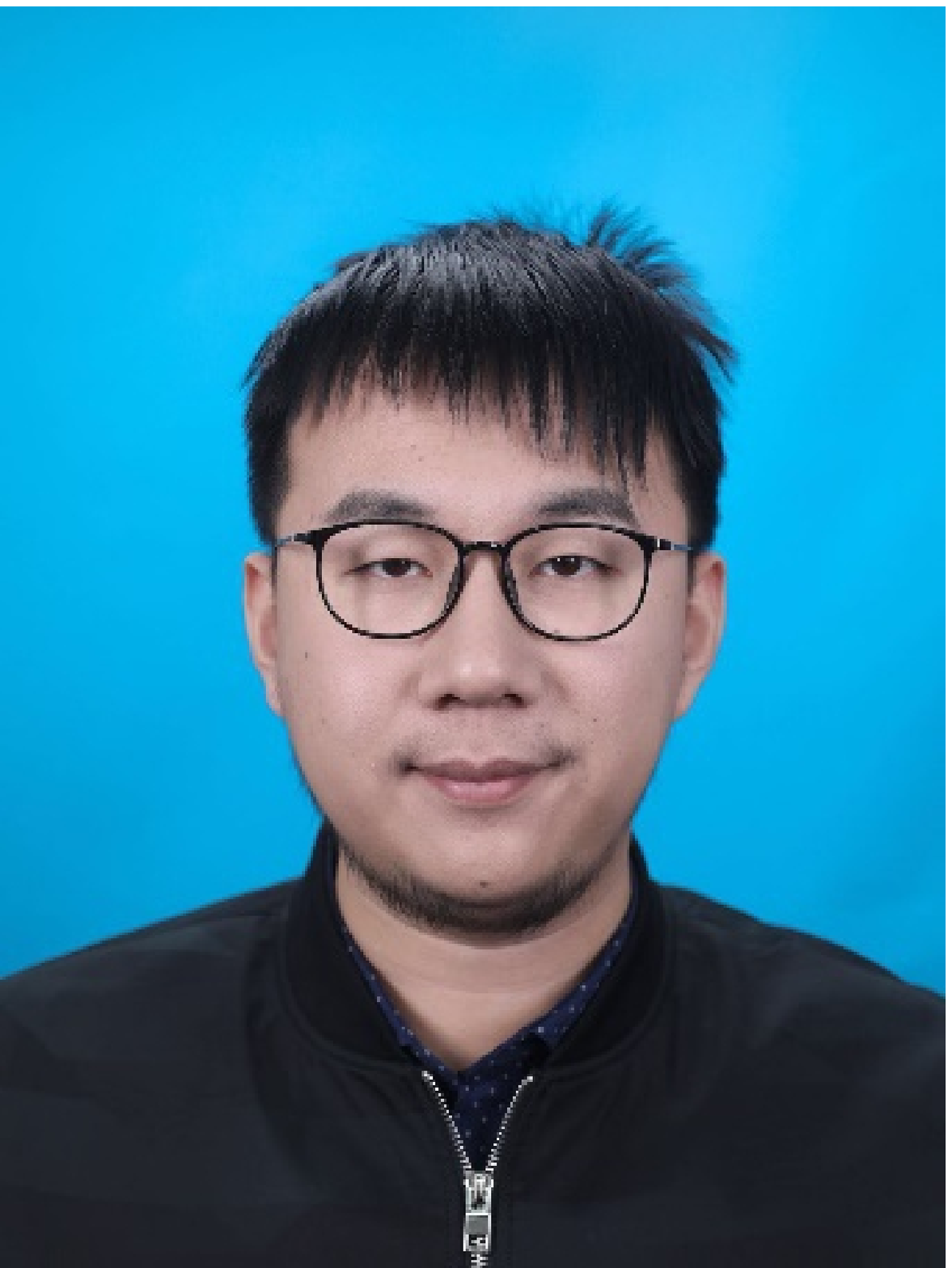}}]
{Xihan Chen}
(S'18-M'20) received the first B.S. degree in electrical engineering from the Beijing University of Posts and Telecommunications, Beijing, China, in 2015, and the second B.S. degree (Hons.) in electrical engineering from the Queen Mary University of London, London, U.K., in 2015. He is currently pursuing the Ph.D. degree with the College of Information Science and Electronic Engineering, Zhejiang University, Hangzhou, China. He was a visiting student with the Department of Electronic and Computer Engineering, University of Toronto, Toronto, ON, Canada, in 2019. His research interests include wireless communication and stochastic optimization.
\end{IEEEbiography}

%\vskip -1\baselineskip plus -1fil

\begin{IEEEbiography}
[{\includegraphics[width=1in,height=1.25in,clip,keepaspectratio]{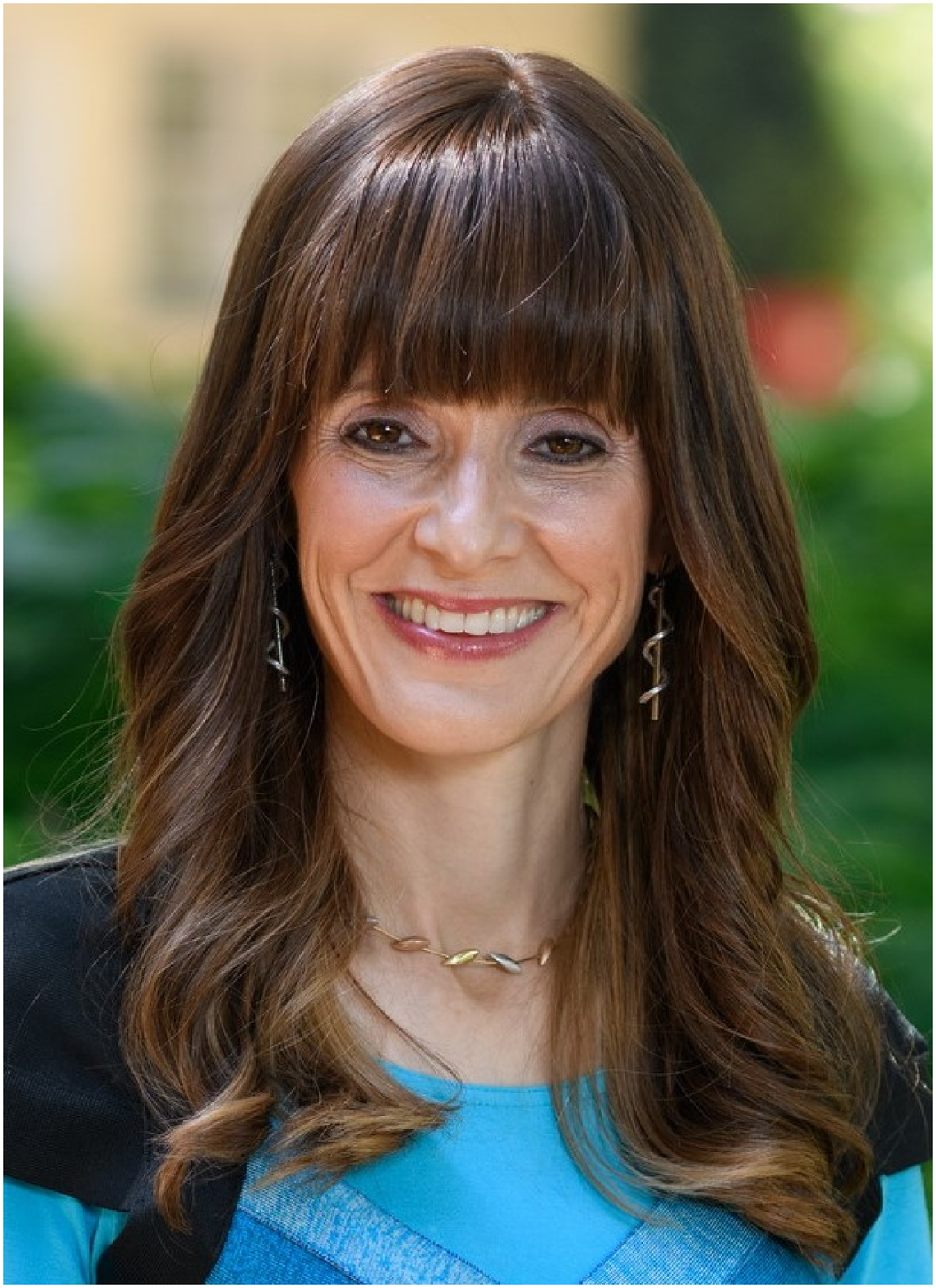}}]
{Yonina C. Eldar} (Fellow, IEEE) received the B.Sc. degree in physics and the B.Sc. degree in electrical engineering both from Tel-Aviv University, Tel-Aviv, Israel, in 1995 and 1996, respectively, and the Ph.D. degree in electrical engineering and computer science from Massachusetts Institute of Technology (MIT), Cambridge, MA, USA, in 2002. She is currently a Professor with the Department of Mathematics and Computer Science, Weizmann Institute of Science, Rehovot, Israel. She was previously a Professor with the Department of Electrical Engineering, Technion, where she held the Edwards Chair in Engineering. She is also a Visiting Professor with MIT, a Visiting Scientist with the Broad Institute, and an Adjunct Professor with Duke University and was a Visiting Professor at Stanford. She is a member of the Israel Academy of Sciences and Humanities (elected 2017) and a EURASIP fellow. Her research interests are in the broad areas of statistical signal processing, sampling theory and compressed sensing, learning and optimization methods, and their applications to biology and optics.

She has received many awards for excellence in research and teaching, including the IEEE Signal Processing Society Technical Achievement Award (2013), the IEEE/AESS Fred Nathanson Memorial Radar Award (2014), and the IEEE Kiyo Tomiyasu Award (2016). She was a Horev Fellow of the Leaders in Science and Technology program at the Technion and an Alon Fellow. She received the Michael Bruno Memorial Award from the Rothschild Foundation, the Weizmann Prize for Exact Sciences, the Wolf Foundation Krill Prize for Excellence in Scientific Research, the Henry Taub Prize for Excellence in Research (twice), the Hershel Rich Innovation Award (three times), the Award for Women with Distinguished Contributions, the Andre and Bella Meyer Lectureship, the Career Development Chair at the Technion, the Muriel \& David Jacknow Award for Excellence in Teaching, and the Technion's Award for Excellence in Teaching (two times). She received several best paper awards and best demo awards together with her research students and colleagues, including the SIAM outstanding Paper Prize, the UFFC Outstanding Paper Award, the Signal Processing Society Best Paper Award and the IET Circuits, Devices and Systems Premium Award, and was selected as one of the 50 most influential women in Israel. She was a member of the Young Israel Academy of Science and Humanities and the Israel Committee for Higher Education. She is the Editor-in-Chief for Foundations and Trends in Signal Processing, a member of the IEEE Sensor Array and Multichannel Technical Committee and serves on several other IEEE committees. In the past, she was a Signal Processing Society Distinguished Lecturer, a member of the IEEE Signal Processing Theory and Methods and Bio Imaging Signal Processing technical committees, and served as an Associate Editor for the IEEE TRANSACTIONS ON SIGNAL PROCESSING, the EURASIP Journal of Signal Processing, the SIAM Journal on Matrix Analysis and Applications, and the SIAM Journal on Imaging Sciences. She was the Co-Chair and Technical Co-Chair of several international conferences and workshops.
\end{IEEEbiography}

\begin{IEEEbiography}
[{\includegraphics[width=1in,height=1.25in,clip,keepaspectratio]{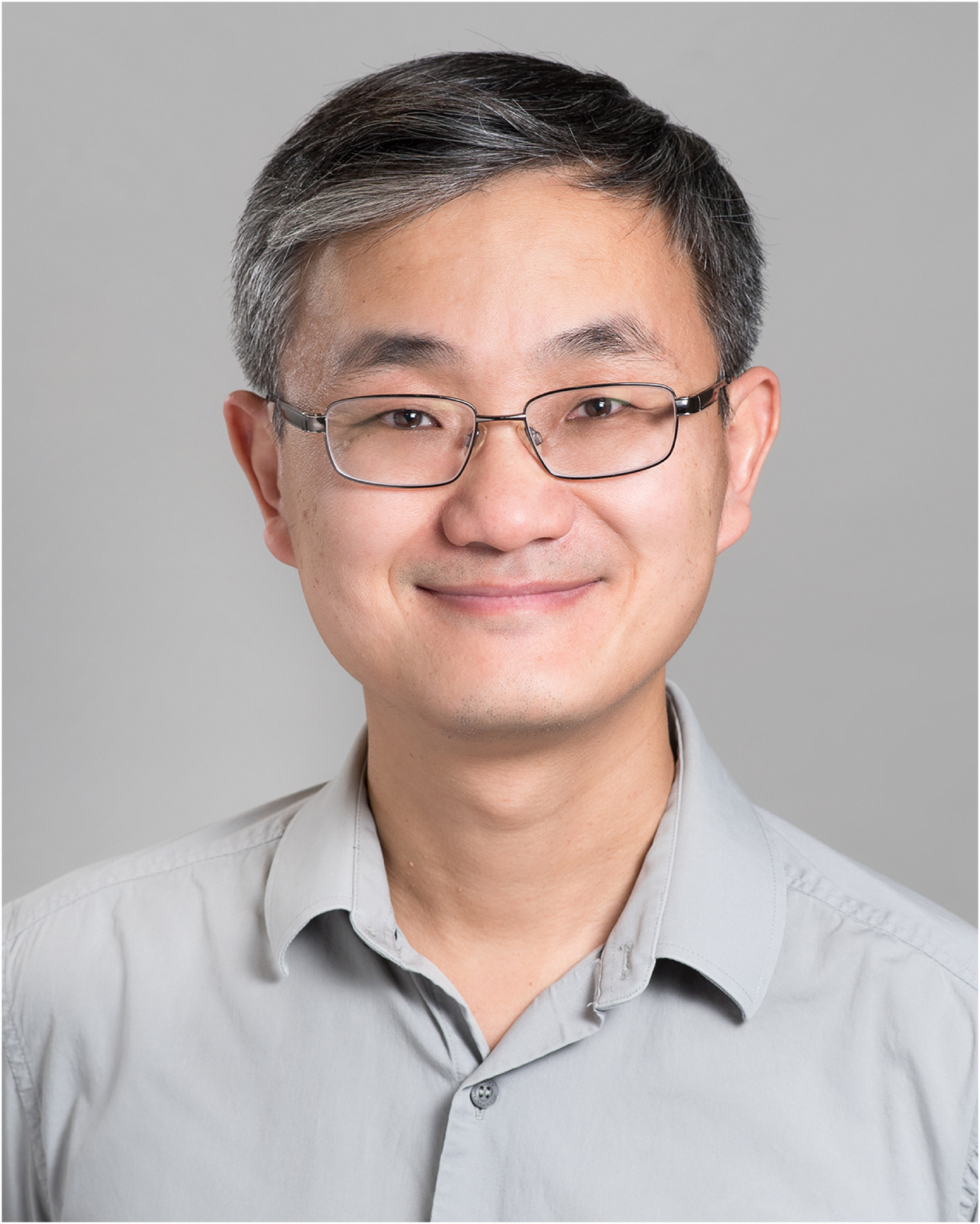}}]
{Wei Yu} (S'97-M'02-SM'08-F'14) 5.received the B.A.Sc. degree in Computer Engineering and Mathematics from the University of Waterloo, Waterloo, Ontario, Canada in 1997 and M.S. and Ph.D. degrees in Electrical Engineering from Stanford University, Stanford, CA, in 1998 and 2002, respectively. Since 2002, he has been with the Electrical and Computer Engineering Department at the University of Toronto, Toronto, Ontario, Canada, where he is now Professor and holds a Canada Research Chair (Tier 1) in Information Theory and Wireless Communications. His main research interests include information theory, optimization, wireless communications, and broadband access networks.

Prof. Wei Yu serves as a Vice President of the IEEE Information Theory Society in 2019-2020, and has served on its Board of Governors since 2015. He is currently an Area Editor for the IEEE Transactions on Wireless Communications, and in the past served as an Associate Editor for IEEE Transactions on Information Theory (2010-2013), as an Editor for IEEE Transactions on Communications (2009-2011), and as an Editor for IEEE Transactions on Wireless Communications (2004-2007). He served as the Chair of the Signal Processing for Communications and Networking Technical Committee of the IEEE Signal Processing Society in 2017-18. Prof. Wei Yu was an IEEE Communications Society Distinguished Lecturer in 2015-16. He received the Steacie Memorial Fellowship in 2015, the IEEE Marconi Prize Paper Award in Wireless Communications in 2019, the IEEE Communications Society Award for Advances in Communication in 2019, the IEEE Signal Processing Society Best Paper Award in 2017 and 2008, the Journal of Communications and Networks Best Paper Award in 2017, the IEEE Communications Society Best Tutorial Paper Award in 2015, an IEEE International Conference on Communications Best Paper Award in 2013, the McCharles Prize for Early Career Research Distinction in 2008, the Early Career Teaching Award from the Faculty of Applied Science and Engineering, University of Toronto in 2007, and an Early Researcher Award from Ontario in 2006. Prof. Wei Yu is a Fellow of the Canadian Academy of Engineering, and a member of the College of New Scholars, Artists and Scientists of the Royal Society of Canada.
\end{IEEEbiography}

\end{document}